%% file: paper.tex
\documentclass[%
prd%
,preprint%
,showpacs%
,superscriptaddress,amssymb
,floatfix]{revtex4}

\usepackage{latexsym}
\usepackage{graphicx}
\usepackage{epstopdf}
\usepackage{bm}

\def\simge{
    \mathrel{\rlap{\raise 0.511ex
        \hbox{$>$}}{\lower 0.511ex \hbox{$\sim$}}}}
\def\simle{
    \mathrel{\rlap{\raise 0.511ex
        \hbox{$<$}}{\lower 0.511ex \hbox{$\sim$}}}}
\def\beqn{\begin{equation}}
\def\eeqn{\end{equation}}
\def\barr{\begin{eqnarray}}
\def\earr{\end{eqnarray}}
\def\bc{\begin{center}}
\def\ec{\end{center}}

\def\Ga    {g_{_{A}}}
\def\Gv    {g_{_{V}}}

\def\Za    {Z_{_A}}
\def\Zv    {Z_{_V}}
\def\rvac {|0\rangle}
\def\lvac {\langle0|}

\def\rproton {| p\rangle}
\def\lproton {\langle p|}
\def\rneutron {| n\rangle}
\def\lneutron {\langle n|}

\newcommand{\Pslash}{p \kern -2mm /}

\newcommand{\Dslash}{\rlap{/}\kern-3.0pt \bm{\mathcal D}}
\newcommand{\QpropPP}[7]{{S_{_{PP}}^{#1}}(#2; #5)^{#4#7}_{#3#6}}
\newcommand{\QpropPW}[7]{{S_{_{PW}}^{#1}}(#2; #5)^{#4#7}_{#3#6}}
\newcommand{\QpropWW}[7]{{S_{_{WW}}^{#1}}(#2, #5)^{#4#7}_{#3#6}}
\newcommand{\QpropPB}[7]{{S_{_{PB}}^{#1}}(#2; #5)^{#4#7}_{#3#6}}
\newcommand{\ThreePtQ}[8]{{{\cal S}_{_{\Gamma}}^{#1}}(#2, #8, #5)^{#4#7}_{#3#6}}
\newcommand{\epsd}[1]{\varepsilon_{#1}}
\newcommand{\nucopD}[1]{(C\gamma_5)_{#1}}
\newcommand{\diq}[9]{\bm{\mathcal #1}_{#3#7}(#2; #6)_{#4#8}^{#5#9}}
\newcommand{\Src}[8]{\bm{ #1}_{#2}(#3; #6)^{#5#8}_{#4#7}}
\newcommand{\SrcAst}[8]{\bm{ #1}^{\ast}_{#2}(#3; #6)^{#5#8}_{#4#7}}

\newcounter{Outline}
\newcounter{Intro}
\newcounter{Analytic}
\newcounter{Numerical}
\newcounter{Conclusions}
\newcounter{Acknowledgments}
\newcounter{Appendix}
\newcounter{Tables}
\newcounter{Figures}
\begin{document}
\bibliographystyle{apsrev}
\preprint{RBRC-323, KEK-TH-869}

\title{Nucleon axial charge from quenched lattice QCD with domain wall fermions}

\author{S. Sasaki}
\affiliation{Department of Physics, University of Tokyo, 
Tokyo 113-0033, Japan}
\author{K. Orginos}
\affiliation{RIKEN BNL Research Center, Brookhaven National Laboratory,
             Upton, NY 11973}
\author{S. Ohta}
\affiliation{Institute of Particle and Nuclear Studies, KEK, Ibaraki 305-0801, Japan}
\affiliation{RIKEN BNL Research Center, Brookhaven National Laboratory,
             Upton, NY 11973}
\author{T. Blum}
\affiliation{RIKEN BNL Research Center, Brookhaven National Laboratory,
             Upton, NY 11973}

\collaboration{RIKEN-BNL-Columbia-KEK Collaboration}


\date{June 5, 2003}

\begin{abstract}
We present a quenched lattice calculation of the nucleon isovector
vector and axial-vector charges $\Gv$ and $\Ga$.  The chiral symmetry
of domain wall fermions makes the calculation of the nucleon axial
charge particularly easy since the Ward-Takahashi identity requires
the vector and axial-vector currents to have the same renormalization,
up to lattice spacing errors of order ${\cal O}(a^2)$. The DBW2 gauge
action provides enhancement of the good chiral symmetry properties of
domain wall fermions at larger lattice spacing than the conventional
Wilson gauge action.  Taking advantage of these methods and performing
a high statistics simulation, we find a significant finite volume
effect between the nucleon axial charges calculated on lattices with
\((1.2\, {\rm fm})^3\) and \((2.4\, {\rm fm})^3\) volumes ($a\approx
0.15$ fm).  On the large volume we find $\Ga = 1.212\pm 0.027({\rm
stat})\pm 0.024({\rm norm})$. The quoted systematic error is the
dominant (known) one, corresponding to current renormalization. We discuss
other possible remaining sources of error. This theoretical first
principles calculation, which does not yet include isospin breaking
effects, yields a value of $\Ga$ only a little bit below the experimental
one, $1.2670 \pm 0.0030$.
\end{abstract}

\maketitle

\newpage


\section{Introduction}
\label{sec:intro}

\ifnum\theIntro=1
\input{text_sections/intro.tex}

\fi


\section{General Analytic Framework}
\label{sec:analytic}

\ifnum\theAnalytic=1
\input{text_sections/analytic.tex}
\fi


\section{Numerical Results}
\label{sec:numeical}

\ifnum\theNumerical=1
\input{text_sections/numerical.tex}
\fi


\section{Conclusions}
\label{sec:conclusions}

\ifnum\theConclusions=1
\input{text_sections/conclusions.tex}
\fi


\ifnum\theAcknowledgments=1
\section*{Acknowledgments}

We thank Roger Horsley for his private communication providing the clover results prior to publication.  We also thank the other members, current and past, of the RIKEN-BNL-Columbia (RBC) collaboration.  In particular, we thank Yasumichi Aoki for his help in optimizing the box source size, Norman Christ for his careful reading of the manuscript, and Chris Dawson for helpful discussion on \(O(a^2, m_fa^2)\) corrections in local current renormalization.  Thanks are also due to RIKEN, Brookhaven National Laboratory and the U.S.\  Department of Energy for providing the facilities essential for the completion of this work.   The numerical calculations were done on the 600 Gflops QCDSP computer at the RIKEN-BNL Research Center.  S. S.\ thanks the JSPS for a Grant-in-Aid for Encouragement of Young Scientists (No. 13740146).  

\fi

\appendix
\ifnum\theAppendix=1
\input{text_sections/appendix.tex}
\fi


\input{ref.tex}
\pagebreak

\ifnum\theTables=1
\input{tab/tab.tex}

\fi


\ifnum\theFigures=1
\input{fig/fig.tex}
\fi

\end{document}

%% file: text_sections/intro.tex
%
%

The axial charge $\Ga$ of the nucleon, or more precisely its ratio to
the vector charge, $\Gv$, appears to be a good test of our understanding
of the structure of the nucleon.  First of all, it is very accurately measured
from neutron $\beta$ decay, $\Ga/\Gv=1.2670\pm0.0030$~\cite{Hagiwara:2002fs}
\footnote{
Note that the Particle Data Group defines \(\Ga\) to be negative because no assumption about the structure of the weak interaction is made.  In this article, assuming the \(V-A\) structure of the weak interaction, the axial form factor in Eq.~\ref{Eq:AxialPart} is defined to make \(\Ga\) positive.}.
And, among the nucleon form factors or moments of structure functions, it is
technically the simplest from the point of view of a lattice QCD numerical
calculation.

Four form factors appear in neutron $\beta$ decay: 
the vector and induced tensor form factors from the vector current,
%
%
\begin{equation}
\langle p| V^+_\mu(x) | n \rangle
= \bar{u}_p [\gamma_\mu g_{_V}(q^2)
             -q_{\lambda} \sigma_{\lambda \mu} g_{_T}(q^2) ] u_n e^{-iq\cdot x},
\end{equation}
and the axial-vector and induced pseudo-scalar form factors from the
axial-vector current,
%
%
\begin{equation}
\langle p| A^+_\mu(x) | n \rangle
= \bar{u}_p 
             [\gamma_\mu  \gamma_5 g_{_A}(q^2)
             -i q_\mu \gamma_5 g_{_P}(q^2) ]  u_n e^{-iq\cdot x}.
\label{Eq:AxialPart}
\end{equation}
Here $q=p_n-p_p$ is the momentum transfer between the proton ($p$)
and neutron ($n$).  In the limit $|{\vec q}| \rightarrow 0$, 
the momentum transfer should
be small because the mass difference of the neutron and proton is only 
about 1.3 MeV.  
This makes the limit $q^2 \rightarrow 0$, where
the vector and axial-vector form factors dominate, a good
approximation.  Their values in this limit are called the vector and
axial charges of the nucleon: $g_{_V} = g_{_V}(q^2=0)$ and
$g_{_A} = g_{_A}(q^2=0)$.  
Experimentally, $g_{_V} = \cos \theta_C$
 (with the Cabibbo mixing angle
$\theta_C$), and $g_{_A} = 1.2670(30) \times g_{_V}$. 

Since they are defined at zero momentum transfer, a naive expectation is that
$\Gv$ and $\Ga$ are easier to calculate on the lattice than 
form factors
which require 
non-zero momentum transfer.
Despite this, quenched QCD lattice calculations with Wilson fermions at finite
lattice cutoff ($a^{-1}\sim 2$ GeV) have underestimated $\Ga$ by about 20\%
~\cite{{Fukugita:1995fh}, {Liu:1994ab}, {Gockeler:1996wg}} (see
Table~\ref{tab:ga_lattice_hist} for a summary of previous calculations).
This suggests systematic errors, which may arise from (1) the quenched approximation, 
(2) operator renormalization,
(3) non-zero-lattice-spacing $a$ and loss of chiral symmetry for Wilson and Kogut-Susskind fermions, and 
(4) finite volume, 
remain in the lattice calculation. 

The first three errors have been addressed in previous calculations.
The SESAM and LHPC collaborations found that unquenching does not
solve the problem as the estimated value $\Ga$ decreases by
5-10\%~\cite{{Gusken:1999as},{Dolgov:2002zm}}.  On the other hand, reducing
the lattice spacing error seems to increase the value, but only by a
small amount, $\simle$
5\%~\cite{{Capitani:1999zd},{Horsley:2000pz}}. Perhaps more important
is the calculation of the renormalization factor $\Za$ for the axial
current.  The one-loop perturbative renormalization factor, used in
the case of Wilson fermions~\cite{{Fukugita:1995fh}, {Liu:1994ab},
{Gockeler:1996wg}, {Dolgov:2002zm},{Gusken:1999as}}, was probably
overestimated.  The QCDSF-UKQCD collaboration reported that the
non-perturbatively calculated renormalization factor ($\Za^{\rm
nonpert}\sim0.8$) is roughly 10\% smaller than the one-loop one
($\Za^{\rm pert}\sim0.9$) in the case of the non-perturbatively ${\cal
O}(a)$ improved Wilson fermions~\cite{Horsley:2000pz} at $a^{-1}\sim
2-3$ GeV.  Thus, the systematic error in the determination of the
renormalization factor appears to be more important than the first two
effects mentioned.
The first two 
systematic errors listed above likely cannot resolve the issue that previous
lattice calculations of $\Ga$ underestimate the experimental value.

The loss of chiral symmetry
on the lattice is potentially significant. As is well known,
$g_{_A}/g_{_V}=1$ in the absence of chiral symmetry breaking in QCD.
Further, in the realistic case of spontaneously broken chiral
symmetry, the ratio is still constrained by the axial Ward-Takahashi
identity; $\partial_{\mu}A^{a}_{\mu}(x)=2mP^{a}(x)$. 
The Goldberger-Treiman relation derives from the nucleon matrix elements 
of the currents on both sides of this identity in the soft pion limit~\cite{Goldberger:1958tr}. 
We can easily understand the deviation of the ratio from unity in the context of the
Gell-Mann-Oakes-Renner relation~\cite{Gell-Mann:1968rz} which is also
related to the axial Ward-Takahashi identity. 
Thus, the explicit breaking of chiral symmetry at non-zero lattice spacing 
$a$ for Wilson fermions may induce significant errors which are only removed 
in the continuum limit.

In this work we use domain wall fermions (DWF), a fermion discretization
scheme with almost perfectly preserved chiral
symmetry~\cite{{Kaplan:1992bt},{Shamir:1993zy},{Narayanan:1993wx}}.  This scheme
introduces a fictitious fifth dimension in addition to the four dimensions of
space-time.  In the limit where the fifth-dimensional extent
\(L_s\) is taken to \(\infty\), DWF preserve the axial Ward Takahashi 
identity~\cite{Furman:1995ky} at non-zero lattice spacing.  With finite \(L_s\) the 
suppression of explicit chiral symmetry breaking is effectively exponential in
quenched simulations if 
the gauge field is sufficiently smooth~\cite{Kikukawa:1999dk,Hernandez:1998et,Hernandez:2000iw,Blum:2000kn,AliKhan:2000iv,Aoki:2002vt}.
This is always true if the
lattice spacing is sufficiently small. In low energy cases like the one
investigated here, the small breaking of the symmetry at finite
\(L_s\) is parametrized by a single universal ``residual mass'' parameter,
\(m_{\rm res}\), acting as an additive quark mass and which is 
defined from the axial Ward-Takahashi
identity~\cite{{Blum:1998ud},{Blum:2000kn}}.  
Furthermore, the DWF scheme
greatly simplifies the non-perturbative determination of the
renormalization of quark bilinear currents~\cite{Blum:2001sr}.  
For example the renormalization factor of local vector and axial-vector
current operators should be equal, $\Za=\Zv$~\cite{Blum:2001sr}.  This means the ratio of the nucleon axial
and vector charges calculated on the lattice directly 
yields the continuum value, ${\it i.e.}$ it is not 
renormalized~\cite{{Blum:2000cf},{Blum:2000cb}}. By employing the DWF scheme,
the ambiguity in the renormalization of quark currents which may be
present and problematic in other fermion discretization schemes is
eliminated.  We emphasize that the DWF calculation of the nucleon
axial charge should not suffer from the systematic errors due to the 
operator renormalization and 
loss of chiral symmetry~\cite{{Blum:2000cf},{Blum:2000cb}}.

However, as is described in more detail in section III, in our
first DWF calculation with the single-plaquette Wilson gauge action at
$\beta = 6.0$ and lattice volume $16^3\times 32 \times 16$ (which
correspond to $a^{-1}\approx 2$ GeV and spacial volume $\sim ({\rm 1.6\, fm})^3$), we
found that $\Ga$ exhibits a fairly strong dependence on the quark
mass~\cite{Blum:2000cb}.  A simple linear extrapolation of $\Ga$ to
the chiral limit yielded a value that was almost a factor of two
smaller than the experiment~\cite{Blum:2000cb}.  
This implied the presence of a large
finite volume effect.  To our surprise, we found no systematic study of
such an effect in the literature. Note also that there is no volume
dependence in the naive quark model~\cite{Isgur:1979be} nor in the MIT bag
model~\cite{Chodos:1974pn}.  In the former the ratio is determined by a simple
spin-isospin algebra, and in the latter it arises from a simple
overlap integral of the upper and lower component of the bag Dirac
wave function.

To address the finite volume issue we need to have at the same time a
sufficiently high lattice cutoff to preserve chiral symmetry
reasonably well and at least two lattice volumes, preferably ones that
are large compared to the charge radius of the proton.  
The Wilson gauge action will not work for this purpose since the
chiral symmetry of DWF in the quenched case degrades rapidly as lattice spacing 
$a$ increases, 
while the computational cost necessitated by a very large
lattice volume would be prohibitive. Fortunately various
``renormalization-group-inspired'' improved gauge actions preserve the
chiral symmetry of DWF well while not demanding a large
cutoff~\cite{{AliKhan:2000iv},{Aoki:2002vt}}.  Thus both requirements,
chiral symmetry and large physical volume, can be met at reasonable
computational cost. Of the relatively well-established candidates in
this class of improved gauge actions, we choose the ``doubly-blocked
Wilson 2 (DBW2)" action~\cite{Takaishi:1996xj,Aoki:2002vt}.

The rest of this paper is organized as follows: in section II the lattice
method for calculating $\Ga/\Gv$ is described.  In section III the
numerical results obtained for both Wilson and DBW2 actions are described in
detail.  Finally, in section IV we summarize the present work and discuss
future directions.

%% file: text_sections/analytic.tex
%
%

\subsection{The vector and axial charges}

As mentioned in the introduction, four form factors are needed to describe
neutron $\beta$ decay: the vector and induced tensor form factors for
the vector current,
%
%
\begin{equation}
\langle p| V^+_\mu(0) | n \rangle
= \bar{u}_p [\gamma_\mu g_{_V}(q^2)
             -q_{\lambda} \sigma_{\lambda\mu} g_{_T}(q^2) ] u_n,
\end{equation}
and the axial and induced pseudo-scalar for the axial current,
%
%
\begin{equation}
\langle p| A^+_\mu(0) | n \rangle
= \bar{u}_p 
             [\gamma_\mu \gamma_5 g_{_A}(q^2)
             -i q_\mu \gamma_5 g_{_P}(q^2) ] u_n.
\end{equation}
The right hand side of each is the most general form consistent with
Lorentz covariance.  
The momentum transfer \(q=p_n-p_p\) becomes very small
in the forward limit 
because of the small mass difference between the neutron and proton.  
In the limit $q^2\to0$, which we take in this work, the vector and axial
form factors dominate the matrix elements. We are neglecting the mass
difference of the neutron and proton, and hence that of up and down
quarks (we also neglect the electromagnetic mass difference.)
For zero quark mass $m=m_u=m_d=0$ the action is symmetric under global
chiral $SU(2)\times SU(2)$ flavor rotations acting on the quark
fields.  If $m\neq 0$, the symmetry is broken down to the vector (flavor)
$SU(2)$ sub-group, and the associated vector charge,
\(g_{_V}\) is still conserved ($\Gv=1$).  This situation is sometimes called
CVC, conserved vector current.  In the real world even this symmetry is
softly broken by the small mass difference between up and down quarks,
\(m_{u}-m_{d}\).  
The explicit violation of the axial-vector symmetry by non-zero quark
mass is sometimes called PCAC, or partially conserved axial-vector
current.  As is well known the axial $SU(2)$ symmetry is also
spontaneously broken. Thus, the axial charge may in general deviate
from unity, $g_{A}\neq 1$.

If the vector symmetry is preserved, a simple exercise
in $SU(2)$ Lie algebra leads to the following (see Appendix
\ref{sec:appendixA}):
%
%
\beqn
\lproton A_{\mu}^{+}\rneutron=2\lproton A_{\mu}^{3} \rproton \nonumber \\
= \lproton A_{\mu}^{u}\rproton-\lproton A_{\mu}^{d} \rproton
\label{eq:axial_matrix}
\eeqn
where \(A_{\mu}^{+} = {\bar u} \gamma_{\mu}\gamma_{5} d\), \(A_{\mu}^{3}=
\frac{1}{2}(A_{\mu}^{u}-A_{\mu}^{d})\),
$A_{\mu}^{u}={\bar u}\gamma_{\mu}\gamma_{5}u$ and $A_{\mu}^{d}=
{\bar d}\gamma_{\mu}\gamma_{5}d$. 
$u$ and $d$ stand for the up and down quark fields.
A similar relation holds for the vector case,
%
%
\beqn
\lproton V_{\mu}^{+}\rneutron=2\lproton V_{\mu}^{3} \rproton \nonumber \\
= \lproton V_{\mu}^{u}\rproton-\lproton V_{\mu}^{d} \rproton
\label{eq:vector_matrix}
\eeqn
where
\(V_{\mu}^{+} = {\bar u} \gamma_{\mu} d\), 
\(V_{\mu}^{3}=\frac{1}{2}(V_{\mu}^{u}-V_{\mu}^{d})\),
$V_{\mu}^{u}={\bar u}\gamma_{\mu}u$ and $V_{\mu}^{d}= {\bar
d}\gamma_{\mu}d$.  The isovector vector charge $\Gv$ and the isovector axial
charge $\Ga$ are defined by the strength of the right-hand side of Eqs. \ref{eq:axial_matrix} and \ref{eq:vector_matrix} in the forward limit ($q^2 \rightarrow 0$).
In addition, the polarized quark distributions in the proton for each flavor $f$,
$\Delta \psi_{f}$, are defined by
the forward matrix element of the flavor axial-vector currents $A_{\mu}^{f}$:
%
%
%
%
%
\beqn
\left\langle k, s\left|\bar \psi_{f} \gamma_\mu \gamma_5 \psi_{f}  \right|k, s\right\rangle
= \Delta \psi_f{\bar u}_{p}(k,s)\gamma_{\mu}\gamma_{5}u_{p}(k,s) 
\eeqn
where $k$ and $s$ are proton four momentum and polarization.
From CVC we find the relation $\Ga= \Delta u - \Delta d$.  

Now consider the conserved electromagnetic current $j_{\mu}^{em}$ expressed in
terms of the flavor vector currents $V^{f}_{\mu}$:
%
%
\beqn
j_{\mu}^{em}=\sum_{f} Q_{f}V_{\mu}^{f}
=
\frac{2}{3}V_{\mu}^{u}-\frac{1}{3}V_{\mu}^{d}+\cdots\;\;.
\eeqn
Here $Q_{f}$ denotes the charge (in units of proton charge $e$) for a
quark of flavor $f$, and the ellipsis denote possible flavors of
heavier quarks which we henceforth ignore.  Since the corresponding
electromagnetic $U(1)$ gauge symmetry assures conservation of electric
charge, for the neutron we find
%
%
\beqn
\lim_{q^2 \rightarrow 0} \lneutron j_{\mu}^{em} \rneutron = 0.
\eeqn
It follows that
%
\beqn
\lim_{q^2 \rightarrow 0} \langle n | V_{\mu}^{d} | n \rangle = 2 \times \lim_{q^2 \rightarrow 0} \langle n |  V_{\mu}^{u} | n \rangle.
\eeqn
On the other hand, under the assumption of CVC we have the following:
%
%
\barr
\lneutron V_{\mu}^{u} \rneutron &=& \lproton V_{\mu}^{d} \rproton \\
\lneutron V_{\mu}^{d} \rneutron &=& \lproton V_{\mu}^{u} \rproton .
\earr
Thus we reach the following relation:
%
%
\beqn
\lim_{q^2 \rightarrow 0}  \langle p | j_{\mu}^{em} | p \rangle = 
\lim_{q^2 \rightarrow 0} \langle p | V_{\mu}^{d} | p \rangle  = 
\lim_{q^2 \rightarrow 0}  \langle p | V_{\mu}^{u}-V_{\mu}^{d} | p \rangle . 
\eeqn
Likewise, it follows that the vector charge, $\Gv$, must be unity (in units of
$\cos \theta_C$ and $e$) under CVC since the proton electric charge is
unity.  As already mentioned, we expect a very small breaking from CVC because
of the physical up and down quark mass difference.  In the axial case,
we expect non-conservation of $\Ga$ due to the small but non-zero up and down
quark masses as well as the spontaneous breakdown of 
chiral symmetry. 

\subsection{Nucleon matrix elements}
\input{text_sections/outline/outline_IIb.tex}
\label{subsec:nucleon_me}

In this subsection we describe our method of lattice numerical calculation of
the axial and vector charges of the nucleon.  Hadronic matrix
elements calculated on the lattice are determined from ratios of the relevant
three-point to two-point correlation functions. Since the charges are defined
at zero-momentum transfer, we do not have to introduce non-zero momentum
projection for the nucleon source and sink for these correlation functions,
nor for the current insertion.  On the other hand since we are dealing with a
spin-1/2 baryon, both correlation functions possess non-trivial Dirac spinor
structure, so appropriate projections are necessary.

The zero-momentum two-point function for the nucleon is given by the
sum over all spatial coordinates, $\vec x$,
%
%
\beqn
\langle {\cal N}(t){\overline {\cal N}}(0)\rangle_{\alpha \beta}
 =\sum_{\vec{x}}\lvac {\cal N}_{\alpha}({\vec x}, t) 
{\overline {\cal N}}_{\beta}({\vec 0},0)\rvac
\label{eq:2ptNucl}
\eeqn
where ${\cal N}(\vec{x}, t)$ can be any operator with the same quantum
numbers as the nucleon, namely unit baryon number, 
$J^{P} = (1/2)^{+}$, and isospin doublet. 
$\alpha$ and $\beta$ denote
Dirac indices. Color and flavor indices are suppressed in the
following unless noted otherwise. The two-point correlation has the
asymptotic form 
%
%
\beqn
\langle {\cal N}(t){\overline {\cal N}}(0)\rangle_{\alpha \beta}
=
\frac{A_{_N}}{2}[1+{\rm sgn}(t)
\gamma_{4}]_{\alpha \beta}e^{-M_{N}|t|}
\eeqn
at large Euclidean time, $t$.  Here $M_{N}$ denotes the ground state mass
of the nucleon. The amplitude $A_{_N}$ is defined as 
$\langle 0|{\cal N}(0)|N\rangle = \sqrt{A_{_N}}u_{N}$. 
In general, the baryon two-point function 
receives contribution from both positive and 
negative-parity states. 
By taking the trace with a projection operator
$P_{+}=(1+\gamma_4)/2$,  
we eliminate contributions from the opposite-parity state in forward time direction.
Details of the parity projection are described in \cite{Sasaki:2001nf}.
Let us abbreviate the notation for the two-point function of the particle
contribution from the desired (positive-parity) state as
%
%
\beqn
G_{_{N}}(t)=\frac{1}{4}{\rm Tr} [P_{+}\langle {\cal N}(t)
{\overline {\cal N}}(0)\rangle].
\eeqn
The factor of 1/4 is our choice of normalization.  At large $t$ this
asymptotically approaches a simple exponential,
%
%
\beqn
G_{_{N}}(t) \sim \exp(-M_{N}t).
\eeqn
For the proton, a standard choice for the interpolating operator is
%
%
\beqn
{\cal N}(x)=\varepsilon_{a b c}[u_a^{T}(x)C\gamma_5d_b(x)]u_{c}(x)
\eeqn
%
where $C$ is the charge conjugation matrix defined as $C=\gamma_{4}\gamma_{2}$,
$abc$ the color indices, $u$ and $d$ the up and down quark fields.

Next, let us define the zero-momentum three-point correlation
function for quark bilinears, ${{\cal O}_{\Gamma}^{(f)}}(x)={\bar
\psi}_{_f}(x)\Gamma
\psi_{_{f}}(x)$:
%
%
\beqn
\langle {\cal N}(t){{\cal O}_{\Gamma}}(t'){\overline {\cal N}}(0)
\rangle_{\alpha \beta}=\sum_{\vec{x}} \sum_{\vec{x}'}
\lvac T\{ {\cal N}_{\alpha}({\vec x}, t){\cal O}_{\Gamma}({\vec x}', t') 
{\overline {\cal N}}_{\beta}({\vec 0},0) \}\rvac
\label{eq:3ptNucl}
\eeqn
where $\Gamma$ is any of the sixteen possible matrices in the Clifford algebra
defined by the Dirac gamma matrices.  When
$t\gg t'\gg 0$, the particle contribution of the zero-momentum
three-point function becomes
%
%
\beqn
\langle {\cal N}(t){{\cal O}_{\Gamma}}(t'){\overline {\cal N}}(0)
\rangle_{\alpha \beta}
\rightarrow
A_{_N}\lim_{q^2\rightarrow 0}g_{_{\Gamma}}(q^2)
\exp(-M_{N}t)
(P_{+}\Gamma P_{+})_{\alpha \beta}.
\eeqn
Note two important points: first, the three-point function vanishes
for $\Gamma$ other than 1,
\(\gamma_{4}\), \(\gamma_{i}\gamma_{5}\) (\(i=1\), 2, 3), and \(\sigma_{ij}\)
(\(i, j=1\), 2, 3) because $P_{+}\Gamma P_{+}=0$ for \(\Gamma\)'s that do not
commute with \(\gamma_4\).  Second, the r.h.s.\ of the above asymptotic
formula does not depend on the insertion point of the operator ${\cal
O}_{\Gamma}$.  Any $t'$-dependence arises from excited state contamination,
{\it i.e.} away from the asymptotic regime.

In this paper, we calculate the isovector (quark-flavored) vector charge
$\Gv$ and the isovector axial charge $\Ga$ of the nucleon.  We define the
spin projected three-point function for the relevant components of the
vector current $V_{\mu}^{f}={\bar \psi}_{f}\gamma_{\mu}{\psi}_{f}$ and the
axial current $A_{\mu}^{f}={\bar
\psi}_{f}\gamma_{\mu}\gamma_{5}{\psi}_{f}$ by taking traces with the
projection operators $P_{\Gamma}=P_{+}\Gamma^{-1}$:
%
%
\barr
G_{V}^{f}(t,t')&=&\frac{1}{4}
{\rm Tr}[P_{V}\langle {\cal N}(t)V_{4}^{f}(t')
{\overline {\cal N}}(0)\rangle] 
\label{eq:3ptProjVec}
\\
G_{A}^{f}(t,t')&=&\frac{1}{4}
{\rm Tr}[P_{A_{i}}\langle {\cal N}(t)A_{i}^{f}(t')
{\overline {\cal N}}(0)\rangle] 
\label{eq:3ptProjAxial}
\earr
where $P_{V}=P_{+}$ and $P_{A_{i}}=P_{+}\gamma_{i}\gamma_{5}$ (\(i=1\), 2, 3).
In order to extract $g_{_{\Gamma}}$ ($\Gamma$ is either $V$ or $A$) on the
lattice, we have to identify a plateau in the ratio of the three- and
two-point functions,
%
%
\beqn
g_{_{\Gamma}}^{\rm lattice}={{G_{_{\Gamma}}^{u}(t,t')-G_{_{\Gamma}}^{d}(t,t')} 
\over {G_{_{N}}(t)}} 
\label{eq:latticeGvGa}
\eeqn
%
in the range of $t>t'$ with fixed $t=t_{\rm sink}-t_{\rm source}$. 

In general, lattice operators receive finite renormalizations relative
to their continuum counterparts since the exact symmetries of the continuum
are usually realized only in the continuum limit, $a\to 0$. Thus
%
%
\beqn
g_{_{\Gamma}}^{\rm ren}=Z_{_{\Gamma}}g_{_{\Gamma}}^{\rm lattice}
\eeqn
requires some independent estimation of $Z_{_{\Gamma}}$, the
renormalization of the quark bilinear currents,
%
%
\beqn
[{\bar \psi}\Gamma \psi]^{\rm ren}=Z_{_{\Gamma}}
[{\bar \psi}\Gamma \psi]^{\rm lattice}.
\eeqn
%
Since DWF possess full chiral
symmetry at non-zero lattice spacing, a lattice conserved vector current
${\cal V}_{\mu}$ 
and partially-conserved axial-vector current 
${\cal A}_{\mu}$ 
which receive no lattice renormalization can be defined, namely
$Z_{_{\cal V}}$=$Z_{_{\cal A}}$=1~\cite{Furman:1995ky}. 
However these conserved currents are point-split 
and require sums over the extra 5th dimension of DWF, so they are somewhat
costly to work with in practice.  Alternatively the local currents 
$V_{\mu}$ and $A_{\mu}$ 
which are naive transcriptions of the
continuum operators are easier to deal with but receive a finite
renormalization since they do not correspond to an exact symmetry of the
action. However, the Ward-Takahashi identity satisfied by both types of
currents is enough to ensure that the lattice renormalizations of the local
currents are equal, $\Zv = \Za$, up to terms of order 
${\cal O}(a^2)$ in the chiral limit and neglecting explicit chiral 
symmetry breaking for DWF at finite \(L_s\).
%
%
\beqn
\left(\frac{\Ga}{\Gv}\right)^{\rm ren}=
\left(
\frac{G_{_A}^u(t,t')-G_{_A}^d(t,t')}{G_{_V}^u(t,t')-G_{_V}^d(t,t')}
\right)^{\rm lattice}+{\cal O}(a^2).
\eeqn
Note that the vector charge computed from the local current provides
an independent estimate of \(\Zv\) since the renormalization of an
operator does not depend on any particular matrix element and the
renormalized, or physical, value of $\Gv$ is 1 by CVC.
%
%
\beqn
\Zv={{G_{_{N}}(t)} \over {{G_{_{_V}}^{u}(t,t')-G_{_{_V}}^{d}(t,t')} }}.
\eeqn
Comparison of \(\Zv\) thus obtained to the value of \(\Za\) from 
the 
relation~\cite{Blum:2000kn}, %
%
\beqn
\langle {\cal A}_{_\mu} (t) [\bar{\psi}\gamma_5 \psi](0) \rangle =
Z_{_A} \langle A_\mu (t) [\bar{\psi}\gamma_5 \psi](0) \rangle,
\eeqn
yields an estimate of the ${\cal O}(a^2)$ 
systematic errors arising
from the method described here. These are discussed in section III.

Next we describe the particular interpolating operators, 
or quark sources and sinks, used to calculate lattice correlation functions.
In our earlier work we used so-called wall-wall correlation functions
constructed with quark sources generated from a unit source
at each spatial site on a fixed source time-slice and summed over all
spatial sites at the sink time-slice. 
Since the wall source or sink is gauge variant, we fix to the Coulmob gauge.
Later we switched to wall-point
correlation functions since they yield smaller statistical errors. For three
point functions this approach is implemented with a sequential
source. We discuss both types of correlation functions in turn.

First, we introduce the forward quark
propagators from a wall source to a wall sink and from
a wall source to a point sink, which may be written with the
gauge fixed point-to-point quark propagator $S_{_{PP}}$:
%
%
\barr
\QpropWW{}{t}{\alpha}{a}{t'}{\beta}{b}
&=&\sum_{{\vec x}}
\QpropPW{}{{\vec x}, t}{\alpha}{a}{t'}{\beta}{b},\\
\QpropPW{}{{\vec x}, t}{\alpha}{a}{t'}{\beta}{b}
&=&\sum_{{\vec y}}
\QpropPP{}{{\vec x}, t}{\alpha}{a}{{\vec y}, t'}{\beta}{b}
\earr
where the subscripts and superscripts denote Dirac and color indices,
respectively.  
The quark three-point function
resulting from insertion of the quark bilinear
operator ${\bar \psi}_{f}\Gamma\psi_f$ is defined as
%
%
\begin{eqnarray}
\ThreePtQ{}{t}{\alpha}{a}{t''}{\beta}{b}{t'}
&=&\sum_{{\vec x}, {\vec y}, {\vec z}}
\QpropPP{}{{\vec x},t}{\alpha}{a}{{\vec y}, t'}{\gamma}{c}
(\Gamma)_{\gamma\gamma'}
\QpropPP{}{{\vec y}, t'}{\gamma'}{c}{{\vec z}, t''}{\beta}{b} \nonumber \\
&=&
\sum_{{\vec z}}
\gamma_{5,\gamma\delta'} 
\QpropPW{*}{{\vec x}, t'}{\delta'}{c}{t}{\delta}{a}
\gamma_{5,\delta\alpha}
(\Gamma)_{\gamma \gamma'} 
\QpropPW{}{{\vec x}, t'}{\gamma'}{c}{t''}{\beta}{b}
\end{eqnarray}
where the second line results from 
$\QpropPP{}{{\vec x}, t}{}{}{{\vec y},t'}{}{}=\gamma_5
\QpropPP{\dagger}{{\vec y}, t'}{}{}{{\vec x}, t}{}{}\gamma_5$. 
Thus, ${\cal S}_{_\Gamma}(t, t', t'')$ is constructed by combining
wall-to-point quark propagators generated from two different source time-slices $t$ and
$t''$ at either end of the lattice with the operator inserted in between them.

The two-point function for the nucleon in Eq.(\ref{eq:2ptNucl}) is expressed
in terms of 
quark propagators as
%
%
\barr
\langle{\cal N}(t){\overline {\cal N}}(0)
\rangle_{\alpha \alpha'}
&=&\varepsilon_{a b c}\varepsilon_{a' b' c'}(C\gamma_5)_{\beta \gamma}
\QpropWW{(d)}{t}{\gamma}{c}{0}{\gamma'}{c'}
(C\gamma_5)_{\beta' \gamma' }
\\ \nonumber
&\times&\left[ 
\QpropWW{(u)}{t}{\alpha}{a}{0}{\alpha'}{a'}
\QpropWW{(u)}{t}{\beta}{b}{0}{\beta'}{b'}
+\QpropWW{(u)}{t}{\alpha}{a}{0}{\beta'}{a'}
\QpropWW{(u)}{t}{\beta}{b}{0}{\alpha'}{b'}
\right]
\earr
Following Ref.~\cite{Woloshyn:1989ae}, the three-point
function in Eq.(\ref{eq:3ptNucl}) is easily obtained from the two-point
function by replacing the ordinary quark propagator by the operator inserted
one, ${\cal S}_{_\Gamma}(t,t',0)$. 
Inserting the d and u quark currents, we obtain
%
%
\barr
\langle{\cal N}(t){\cal O}^{(d)}_{\Gamma}(t'){\overline {\cal N}}(0)
\rangle_{\alpha \alpha'}
&=&\varepsilon_{a b c}\varepsilon_{a' b' c'}(C\gamma_5)_{\beta \gamma}
\ThreePtQ{(d)}{t}{\gamma}{c}{0}{\gamma'}{c'}{t'}
(C\gamma_5)_{\beta' \gamma' }
\\ \nonumber
&\times&\left[ 
\QpropWW{(u)}{t}{\alpha}{a}{0}{\alpha'}{a'}
\QpropWW{(u)}{t}{\beta}{b}{0}{\beta'}{b'}
+\QpropWW{(u)}{t}{\alpha}{a}{0}{\beta'}{a'}
\QpropWW{(u)}{t}{\beta}{b}{0}{\alpha'}{b'}
\right]
\earr
and 
%
%
\barr
\langle{\cal N}(t){\cal O}^{(u)}_{\Gamma}(t'){\overline {\cal N}}(0)
\rangle_{\alpha \alpha'}
&=&\varepsilon_{a b c}\varepsilon_{a' b' c'}(C\gamma_5)_{\beta \gamma}
\QpropWW{(d)}{t}{\gamma}{c}{0}{\gamma'}{c'}
(C\gamma_5)_{\beta' \gamma'}
\\ \nonumber
&\times&\left[ 
\ThreePtQ{(u)}{t}{\alpha}{a}{0}{\alpha'}{a'}{t'}
\QpropWW{(u)}{t}{\beta}{b}{0}{\beta'}{b'}
+\ThreePtQ{(u)}{t}{\alpha}{a}{0}{\beta'}{a'}{t'}
\QpropWW{(u)}{t}{\beta}{b}{0}{\alpha'}{b'}\right.\\ \nonumber
&&\left.+\QpropWW{(u)}{t}{\alpha}{a}{0}{\alpha'}{a'}
\ThreePtQ{(u)}{t}{\beta}{b}{0}{\beta'}{b'}{t'}
+\QpropWW{(u)}{t}{\alpha}{a}{0}{\beta'}{a'}
\ThreePtQ{(u)}{t}{\beta}{b}{0}{\alpha'}{b'}{t'}\right].
\earr 
The nucleon three-point function is the sum of the up and down quark
contributions.  
The spin projected three-point functions are  obtained 
from Eqs.\ref{eq:3ptProjVec} and \ref{eq:3ptProjAxial}.


To enhance the signal, a point sink is more desirable
than an extended sink. The wall-point type of three-point functions 
is implemented using the so-called sequential source method~\cite{Bernard:1985ss,Martinelli:1989rr,Dolgov:2002zm}.
In addition we use box source instead of wall source to enhance the coupling to the ground state:
%
%
\begin{equation}
\QpropPB{}{{\vec x}, t}{\alpha}{a}{t'}{\beta}{b}
=\sum_{0\le{\vec y}\le B}
\QpropPP{}{{\vec x}, t}{\alpha}{a}{{\vec y}, t'}{\beta}{b}
\end{equation}
We adjust the box size \(B\) to about 1 fm.
In describing the construction of the sequential source, 
it is convenient to introduce  the ``diquark" propagators:
%
%
%
%
%
\begin{eqnarray}
\diq{D}{{\vec y},t}{\alpha}{\beta'}{a'}{0}{\alpha'}{\beta}{a} &=& 
\epsd{abc}\epsd{a'b'c'}
\nucopD{\delta\beta}\nucopD{\delta' \beta'}\nonumber\\
&\times&\left[ 
               \QpropPB{(u)}{{\vec y},t}{\beta}{b}{0}{\beta'}{b'}
               \QpropPB{(u)}{{\vec y},t}{\gamma}{c}{0}{\gamma'}{c'}
             +\QpropPB{(u)}{{\vec y},t}{\gamma}{b}{0}{\beta'}{b'}
               \QpropPB{(u)}{{\vec y},t}{\beta}{c}{0}{\gamma'}{c'}    \right] 
\label{eq:diq_d}
\end{eqnarray} 
and
%
%
\begin{eqnarray}
\diq{U}{{\vec y},t}{\alpha}{\beta'}{b'}{0}{\alpha'}{\beta}{b} 
&=&\epsd{abc}\epsd{a'b'c'}
\QpropPB{(d)}{{\vec y},t}{\gamma}{c}{0}{\gamma'}{c'}
\\
&\times&
\left[\nucopD{\beta \gamma}\nucopD{\beta' \gamma'}     
  \QpropPB{(u)}{{\vec y},t}{\alpha}{a}{0}{\alpha'}{a'} 
+\nucopD{\delta \gamma}\nucopD{\delta' \gamma'}
  \QpropPB{(u)}{{\vec y},t}{\delta}{a}{0}{\delta'}{a'}
  \delta_{\alpha \beta}\delta_{\alpha' \beta'} \right.\nonumber \\
&+&\left.
  \nucopD{\beta \gamma}\nucopD{\delta' \gamma'}
  \QpropPB{(u)}{{\vec y},t}{\alpha}{a}{0}{\delta'}{a'}
  \delta_{\alpha' \beta'}         
+\nucopD{\delta \gamma}\nucopD{\beta' \gamma'}  
  \QpropPB{(u)}{{\vec y},t}{\delta}{a}{0}{\alpha'}{a'} 
  \delta_{\alpha \beta} \right].\nonumber
\label{eq:diq_u}
\end{eqnarray} 
The ``down diquark'' ($\cal D$) and  ``up diquark'' ($\cal U$)  are defined by 
the down quark removed propagator and one up quark removed propagator
from the nucleon two-point function. Now using the diquark we can reconstruct
the point-to-wall type of the nucleon two-point function as
%
%
\beqn
\langle{\cal N}(t){\overline {\cal N}}(0)
\rangle_{\alpha \alpha'}
=\sum_{\vec y}
\diq{D}{{\vec y},t}{\alpha}{\beta'}{b}{0}{\alpha'}{\beta}{a} 
\QpropPB{(d)}{{\vec y},t}{\beta}{a}{0}{\beta'}{b}
=\frac{1}{2}\sum_{\vec y}
\diq{U}{{\vec y},t}{\alpha}{\beta'}{b}{0}{\alpha'}{\beta}{a} 
\QpropPB{(u)}{{\vec y},t}{\beta}{a}{0}{\beta'}{b}.
\eeqn

In terms of the diquarks the three-point functions of an arbitrary
quark bilinear operator ${\bar \psi}_{f}\Gamma\psi_{f}$
at a location $({\vec z}, t')$ can be written for the down quark
%
%
\beqn
\langle{\cal N}(t){\cal O}^{(d)}_{\Gamma}(t'){\overline {\cal N}}(0)
\rangle_{\alpha \alpha'}=
\sum_{{\vec y}, {\vec z}}
\diq{D}{{\vec y},t}{\alpha}{\beta'}{a'}{0}{\alpha'}{\beta}{a} 
\QpropPP{(d)}{{\vec y},t}{\beta}{a}{{\vec z},t'}{\delta}{e}
(\Gamma)_{\delta \delta'}
\QpropPB{(d)}{{\vec z},t'}{\delta'}{e}{0}{\beta' }{a'}
\eeqn
and for the up quark
%
%
\beqn
\langle{\cal N}(t){\cal O}^{(u)}_{\Gamma}(t'){\overline {\cal N}}(0)
\rangle_{\alpha \alpha'}=
\sum_{{\vec y}, {\vec z}}
\diq{U}{{\vec y},t}{\alpha}{\beta'}{b'}{0}{\alpha'}{\beta}{b} 
\QpropPP{(u)}{{\vec y},t}{\beta}{b}{{\vec z},t'}{\delta}{e}
(\Gamma)_{\delta \delta'}
\QpropPB{(u)}{{\vec z},t'}{\delta'}{e}{0}{\beta' }{b'}.
\eeqn

For the construction of the three point functions we need
the backward propagators from the sink point $({\vec y}, t)$ 
to the operator insertion point $({\vec z}, t')$.
However, it is highly expensive to prepare 
the required point-to-point quark
propagators from all 
points $(\vec y,
t)$.  This difficulty is easily
circumvented by directly computing the generalized quark propagators
$\diq{D}{{\vec y},t}{\alpha}{\null}{\null}{0}{\alpha'}{\null}{\null} 
\QpropPP{(d)}{{\vec y},t}{\null}{\null}{{\vec z},t'}{\null}{\null}$ and
$\diq{U}{{\vec y},t}{\alpha}{\null}{\null}{0}{\alpha'}{\null}{\null} 
\QpropPP{(u)}{{\vec y},t}{\null}{\null}{{\vec z},t'}{\null}{\null}$ 
with the sequential source method.

Before describing details of the sequential source propagator, 
we should apply the spin projection $P_{\Gamma}$
to diquarks in order  to reduce the cost 
from having to calculate all $4 \times 4$ matrices for external spinor  
indices ($\alpha, \alpha'$). In this article, we only need two kinds of 
spin projections, i.e. $P_{V}$ and $P_{A_{3}}$ so that it reduces the amount
of calculations by a factor of eight in comparison with the unprojected case.
The spin projected source for a down quark insertion is
%
%
\begin{equation}
\Src{\Phi}{d}{{\vec y},t}{\alpha}{a}{0}{\beta}{b} = \frac{1}{4}
\diq{D}{{\vec y},t}{\gamma}{\alpha}{a}{0}{\gamma'}{\beta}{b}
(P_{\Gamma})_{\gamma' \gamma},
\label{eq:d_source}
\end{equation}
and for the up quark
\begin{equation}
\Src{\Phi}{u}{{\vec y},t}{\alpha}{a}{0}{\beta}{b} = \frac{1}{4}
\diq{U}{{\vec y},t}{\gamma}{\alpha}{a}{0}{\gamma'}{\beta}{b}
(P_{\Gamma})_{\gamma' \gamma}.
\label{eq:u_source}
\end{equation}

Finally, the sequential source down quark propagator 
is
\begin{equation}
\Src{\Sigma}{d}{t,0}{\alpha}{a}{{\vec z}, t'}{\beta}{b} = \left(
\sum_{{\vec y}, t''}
\gamma_{5, \beta \gamma} 
\QpropPP{(d)}{{\vec z},t'}{\gamma}{b}{{\vec y},t}{\gamma'}{e} 
\gamma_{5, \gamma' \delta} 
\delta_{t'', t}
\SrcAst{\Phi}{d}{{\vec y},t''}{\alpha}{a}{0}{\delta}{e}\right)^*,
\label{eq:seq_d_prop}
\end{equation}
and the sequential source up quark propagator is
\begin{equation}
\Src{\Sigma}{u}{t,0}{\alpha}{a}{{\vec z}, t'}{\beta}{b} = \left(
\sum_{{\vec y}, t''}
\gamma_{5, \beta \gamma} 
\QpropPP{(u)}{{\vec z},t'}{\gamma}{b}{{\vec y},t}{\gamma'}{e} 
\gamma_{5, \gamma' \delta} 
\delta_{t'', t}
\SrcAst{\Phi}{u}{{\vec y},t''}{\alpha}{a}{0}{\delta}{e}\right)^*.
\label{eq:seq_u_prop}
\end{equation}
which may be calculated by solving the matrix equations
%
%
\beqn
\sum_{{\vec x}, t''}
\Src{\Sigma}{f}{t, 0}{\alpha}{a}{{\vec x}, t''}{\beta}{b} \;\;
\bm{\mathcal M}^{\dagger}({\vec x}, t''; {\vec z}, t')_{\beta \gamma}^{b c}
=\delta_{t t'}
\Src{\Phi}{f}{{\vec z},t'}{\alpha}{a}{0}{\gamma}{c}
\eeqn
where 
\(\bm{\mathcal M}\) is the Dirac matrix.
Consequently, in terms of the sequential source propagator, the spin projected three point function for the down quark is
written
\begin{equation}
G_{\Gamma}^{d}(t,t')=\sum_{{\vec z}}
\Src{\Sigma}{d}{t,0}{\alpha}{a}{{\vec z}, t'}{\beta}{b}(\Gamma)_{\beta \gamma}
\QpropPB{(d)}{{\vec z},t'}{\gamma}{b}{0}{\alpha}{a}
\label{eq:proj_nulc_3pt_d}
\end{equation}
and for the up quark is
\begin{equation}
G_{\Gamma}^{u}(t,t')=\sum_{{\vec z}}
\Src{\Sigma}{u}{t,0}{\alpha}{a}{{\vec z}, t'}{\beta}{b}(\Gamma)_{\beta \gamma}
\QpropPB{(u)}{{\vec z},t'}{\gamma}{b}{0}{\alpha}{a}.
\label{eq:proj_nulc_3pt_u}
\end{equation}

In the case of keeping the up and down quark masses equal, the total cost for
computing the sequential source propagator is a factor of two over the cost for
wall-wall correlation functions. However, the resulting 
box-point correlation functions
yield smaller statistical errors.  

%% file: text_sections/outline/outline_IIb.tex
\ifnum\theOutline=1
\noindent \framebox{Begin \ outline \hspace{5.0in} }
\begin{enumerate}
\item Interpolating operator for proton ; \( {\cal N}= \epsilon_{abc} [u_a^T C \gamma_5d_b]u_c\).
\item Time-slice correlation functions (the rest frame, the zero-momentum):\\
\(
\langle {\cal N}(t){\bar {\cal N}}(0) \rangle_{\alpha \beta}
=\sum_{\vec x} \langle T\{ {\cal N}({\vec x}, t){\bar {\cal N}}({\vec 0},0)  \}\rangle_{\alpha \beta}
\)
\\
\(
\langle {\cal N}(t){\cal O}(t'){\bar {\cal N}}(0) \rangle_{\alpha \beta}
=\sum_{\vec x}\sum_{{\vec x}'}  
\langle T\{ {\cal N}({\vec x}, t){\cal O}({\vec x}', t'){\bar {\cal N}}({\vec 0},0)  \}\rangle_{\alpha \beta}
\)
\item Two-point function:\\
\(\displaystyle
G_{_N}(t) = \frac{1}{4}{\rm Tr}\{ P_{+}  \langle {\cal N}(t){\bar {\cal N}}(0) \rangle\},
\)
with the (particle and parity) projection operator \(P_{+}=(1+\gamma_4)/2\).

\item Three-point functions,
\begin{itemize}
\item vector:
\(\displaystyle
G_{_V}^{u,d}(t,t') = \frac{1}{4}{\rm Tr} \{P_{+}
\langle  {\cal N}(t){V_{4}^{u,d}}(t'){\bar {\cal N}}(0) \rangle \},
\)
\item axial:
\(\displaystyle
G_{_A}^{u,d}(t,t') = \frac{1}{4} {\rm Tr} \{
P_{+}\gamma_i\gamma_5 
\langle  {\cal N}(t){A_{i}^{u,d}}(t'){\bar {\cal N}}(0) \rangle \},
\) (\(i=1,2,3\))
\end{itemize}
with fixed \(t'=t_{\rm source}-t_{\rm sink}\) and \(t < t'\).

\item From the lattice estimate
\[
g_{_\Gamma}^{\rm lattice}
= \frac{G_{_\Gamma}^u(t,t')-G_{_\Gamma}^d(t,t')}{G_{_N}(t)},
\]
with \(\Gamma = V\) or \(A\), the continuum value
\[
g_{_\Gamma} = Z_{_\Gamma} g_{_\Gamma}^{\rm lattice},
\]
is obtained.

\item Non-perturbative renormalizations, defined by
\[
[\bar{u} \Gamma d]_{\rm ren} = Z_{_\Gamma} [\bar{u}\Gamma d]_0,
\]
satisfies \(Z_{_A} = Z_{_V}\) well, so that
\[
\left(\frac{g_{_A}}{g_{_V}}\right)^{\rm ren} =
\left(
\frac{G_{_A}^u(t,t')-G_{_A}^d(t,t')}{G_{_V}^u(t,t')-G_{_V}^d(t,t')}
\right)^{\rm lattice}.
\]
\end{enumerate}
\noindent \framebox{End \ outline \hspace{5.0in} }
\vspace{0.25in}
\fi

%% file: text_sections/numerical.tex
%
%
\input{text_sections/outline/outline_III.tex}

We have performed quenched lattice calculations using two different gauge
actions, the standard Wilson and the improved 
DBW2
~\cite{Takaishi:1996xj}.  Details and some relevant results of both
simulations are summarized in
Tables~II,~\ref{tab:simulation_information},
and~\ref{tab:fermionic_scale_param}.  We describe the nucleon matrix element
results for each one separately, then compare them and draw some conclusions.

\subsection{Wilson gauge action results at $\beta = 6.0$}
\input{text_sections/outline/outline_IIIa.tex}
\label{subsec:wilson}

We have performed a quenched simulation on a $16^3 \times 32$ lattice
with the standard single-plaquette Wilson action at $\beta =
6/g^2=6.0$ which corresponds to a lattice cut-off of $a^{-1}=1.922$
GeV set by the $\rho$ mass~\cite{Blum:2000kn}. Quark propagators were
generated with four bare masses, $m_f=$0.02, 0.03, 0.04 and 0.05, using
DWF with $L_s=16$ and $M_5=1.8$. The nucleon matrix elements were averaged on
a set of 400 gauge configurations. Hadron masses computed on these lattices
are tabulated in Table~\ref{tab:hadrons_wilson_b60}.
Preliminary results for the nucleon
charges were first reported in \cite{Blum:2000cb}.

We calculated wall-source quark propagators on each Coulomb-gauge-fixed
configuration for both periodic and anti-periodic boundary conditions in the
time direction for the quarks. A simple linear combination of these
propagators then yields a forward (or backward) in time propagator.  To compute
the correlation functions, we employed the wall-wall method described in the
previous section with source locations fixed
at $t_{src}=5$ and $t^\prime_{src}=21$.

In Figure~\ref{fig:ZvInsertion_W6} we show the dependence of the vector
renormalization, $\Zv=1/\Gv^{\rm lattice}$ on the location of the current
insertion. A
good plateau is observed in the middle region between the source and sink. The
quoted errors are estimated by a single elimination jack-knife method.  The
dashed lines represent the average value and statistical error in the
time-slice range $5
\leq t-t_{\rm src} \leq 11$.  The mass dependence of $\Zv$ is rather mild as
seen in Figure~\ref{fig:Zv_mass_dep_W6} and given in
Table~\ref{tab:results_wilson_b60}.  The values 0.7601(31) for a linear fit
and 0.7610 (52) for a quadratic fit at $m_f=0$ agree well with
$\Za=0.7555(3)$~\cite{Blum:2000kn}, which was obtained from a calculation of
meson two-point correlation functions.  The discrepancy $\Delta Z
\equiv |1-\Za/\Zv|$ is less than 0.6\% which implies the
${\cal O}(a^2)$ error that remains after taking the $m_f\to0$ limit is
quite small.

As is seen in Figure~\ref{fig:LDuLDdInsertion_W6}, plateaus are evident for the spin-dependent distribution functions, $\Delta u$ and $\Delta d$, in the range $5\leq t-t_{\rm src} \leq 11$.  
Thus, we compute the charge ratios
$(\Ga/\Gv)^{\rm lattice}$ at each $m_{f}$ by taking a weighted average
over this time slice range.  In Figure~\ref{fig:gAgV_W6} a strong
dependence on $m_f$ appears.  A simple linear extrapolation to
$m_{f}=0$ yields $0.812 (112)$, which is roughly 2/3 of the
experimental value.  
However, a simple linear ansatz may not describe the data which show increasing downward curvature for lighter quark mass (note that the points are correlated in this quenched calculation since they are computed on the same gauge configurations).
In general chiral logarithms may appear and were
considered. In fact, the data are not compelling for such terms,
arising in either quenched or full chiral perturbation
theory~\cite{Kim:1998bz,Jenkins:1991es}.
The results for each mass are reproduced in Table~\ref{tab:results_wilson_b60}.

This implies the existence of other systematic errors.  As was
mentioned in the introduction, a large systematic error in previous
lattice calculations of $\Ga$ came from the determination of the
renormalization constant $\Za$.  As shown above using DWF, the value
of $\Ga$ is determined in a fully nonperturbative way, with or without
explicit renormalization. The systematic error stemming from the
incomplete cancellation of renormalization factors in the ratio is
less than 1\% as we saw by comparing $\Zv=1/g_{_V}^{\rm lattice}$ and $\Za$ calculated
from meson two-point functions.  In addition, comparing the chirally
extrapolated values of $(\Ga/\Gv)^{\rm lattice}$ and $\Za \Ga^{\rm
lattice}$ leads to an even smaller error, 
although it relies on the linear extrapolation which was not very compelling.  Another possible systematic error is the
contribution of excited states, the presence or absence of which was
checked by slightly enlarging the separation between wall
sources, $t_{\rm src}(=5)$ and $t_{\rm sink}(=27)$. While the larger
separation induces more noise in the signal, the central value of
$\Ga$ is essentially unchanged for each quark mass; thus we cannot
detect a systematic effect outside of the statistical errors. Still,
this source of error appears to be small.

Detailed detection of quenching effects: quenched chiral logarithms,
unsuppressed fermionic zero modes, and the absence of the physical pion
cloud, is beyond our scope at present since these require very light
quark masses and correspondingly large statistics.  Thus, by process
of elimination we are lead to focus on finite volume effects which we
discuss in the next section.  The volume employed for the calculations
in this subsection is roughly $(1.5-1.6\,{\rm fm})^3$ which can barely
accommodate a proton with mean square radius estimated to be about 0.8
fm~\cite{Hagiwara:2002fs}.

\subsection{DBW2 action results at $\beta = 0.87$}
\input{text_sections/outline/outline_IIIb.tex}
\label{subsec:dbw2}

To determine $\Ga$ in a large physical volume, say $\simge (2
{\rm\,fm})^3$, we have performed a DWF simulation on a lattice with
larger spacing. In general, it is difficult to maintain the good
chiral properties of DWF as $a$ increases at fixed $L_s$, especially
with the Wilson gauge action~\cite{Blum:2000kn,AliKhan:2000iv}.
It has been shown
that the Iwasaki gauge action
enables studies of quenched DWF with smaller $L_s$ than the Wilson
gauge action~\cite{Wu:1999cd,AliKhan:2000iv}.
Recent quenched studies by the
RBC collaboration have shown that the chiral symmetry of DWF are even
better with a similar type of renormalization group improved gauge
action, DBW2~\cite{Aoki:2002vt}. 
The chiral symmetry of DWF with DBW2 is significantly
improved over the Iwasaki action. A very small additive
quark mass $m_{\rm res}\sim 0.8$ MeV is achieved on a lattice with
$a^{-1}\approx 1.3$ GeV and $L_s=16$. Good scaling behavior of the light
hadron spectrum is observed as well~\cite{Aoki:2002vt}.

To study finite volume effects numerical simulations were performed at $\beta
=0.87$ ($a\approx 0.15$ fm) on two lattice sizes, $8^3 \times 24$ and $16^3
\times 32$, with $L_{s}=16$ and $M_{5}=1.8$.  Our results are analyzed on 400
quenched gauge configurations for the smaller lattice ($La\sim 1.2\, {\rm fm}$)
and 416 configurations for the larger lattice ($La\sim 2.4\, {\rm fm}$). Hadron
masses computed in this calculation are summarized in Table
\ref{tab:masses_dbw2_b087}.  Meson masses ($m_{\pi}$ and $m_{\rho})$ for the
$16^3 \times 32$ lattice are evaluated from 100 configurations.

In this calculation, we utilize the sequential quark propagator method to
compute three-point functions as described in section II. We checked for
consistency with the wall-wall method on the smaller $8^3 \times 24$
lattice. The sequential- and wall-type quark propagators in Coulomb gauge were
computed at five evenly spaced values of $m_f$ ranging from 0.02 to 0.10.
The smallest quark mass corresponds to a pion mass $m_{\pi} \approx 390$ MeV.
The Nucleon source and sink were separated by about 1.5 fm, which corresponds
to the same physical separation in time used in the calculation with
the Wilson gauge action at $\beta =6.0$.  A preliminary version of the results
presented below was first reported in \cite{{Sasaki:2001th},{Ohta:2002ns}}.

First, we check whether $\Zv=\Za$ is true even on this coarse lattice.  The vector renormalization $1/\Gv^{\rm lattice}$ is plotted against the location of current insertions in Figure~\ref{fig:ZvInsertion_D087}.  The data are calculated on the larger spatial volume with the sequential quark propagator method.  We take a weighted average of $1/\Gv^{\rm lattice}$ with the three middle points ($t-t_{\rm src}=4, 5, 6$) to evaluate the vector renormalization $\Zv$.  The dependence of $\Zv$ on $m_{f}$ is shown in Figure~\ref{fig:ZAV} and given in Table~\ref{tab:results_dbw2_b087_16cub}. 
In general \({\cal V}_{\mu} = \Zv V_{\mu} + {\cal O}(a^2,m_{f} a^2)\) and \({\cal A}_{\mu} = \Za A_{\mu} + {\cal O}(a^2,m_{f} a^2)\), where \({\cal V}_\mu\) and \({\cal A}_\mu\) denote the conserved vector currents.
It is not so apparent in our data.  
A linear extrapolation yields $\Zv = 0.7952 (13)$ at $m_{f}=0$, while a linear plus quadratic extrapolation gives the value 0.7991(25).
The RBC collaboration obtained the renormalization factor of the axial-vector current $\Za$ nonperturbatively from a calculation of meson two-point correlation functions~\cite{{Blum:2000kn},{Aoki:2002vt}}.  It was found $\Za=0.77759(45)$ in the massless limit~\cite{Aoki:2002vt} which is smaller than the value of $\Zv$ obtained above by 2-3\%.  This discrepancy may be caused by an order ${\cal O}(a^2)$ lattice artifact.

To explore this possibility further, we evaluate the renormalization
factor of different vector currents. According to
Sec.\ref{sec:analytic}, the conserved current ${\cal V}_{\mu}$
guarantees that nucleon matrix elements of $j_{\mu}^{em}$, ${\cal
V}_{\mu}^{d}$ and ${\cal V}_{\mu}^{u}-{\cal V}_{\mu}^{d}$ should be
identical.  
The local lattice
currents $V_{\mu}^{f}$ are renormalized as ${\cal V}^{f}_{\mu}=\Zv^{f}
V^{f}_{\mu}+{\cal O}(a^2)=\Zv V^{f}_{\mu}+{\cal O}(a^2)$ in the chiral
limit. Figure~\ref{fig:ZVSYS} shows the values of $\Zv^{f}$ as well as the
value of $\Za$ from~\cite{Aoki:2002vt}. 
The difference among values of $\Zv$ appears
independent of $m_{f}$ within statistical errors, and the discrepancy
between the smallest and the largest is comparable to that between
$\Zv$ and $\Za$ noted above. We also note that the discrepancy is
larger at this lattice spacing, by roughly a factor $(1.922/1.3)^2$,
than the corresponding one for the Wilson gauge action results
discussed earlier. Of course, since the two gauge actions have
different ${\cal O}(a^2)$ errors, the comparison is only a crude
one. We conclude that $\Zv=\Za$ is 
satisfied up to small
discretization errors of ${\cal O}(a^2)$ on this coarse lattice.

In Figure~\ref{fig:BareDeltaq} we plot unrenormalized spin-dependent
densities $\Delta u$ and $\Delta d$, which are calculated with the
sequential source propagator~\cite{Martinelli:1989rr}, 
against the location of the current insertion.  In this calculation, the
sequential source propagator was calculated with a 
box source and a point
sink, so the resulting three-point function has no time reflection symmetry
about the midpoint between $t_{\rm src}$ and $t_{\rm sink}$ because excited
state contamination is worse for the nucleon propagating between the operator
and the point sink.  In Figure~\ref{fig:BareDeltaq} the plateaus appear shifted
toward the wall source, %
as expected.  Next we evaluate the bare value of $\Ga^{\rm
lattice}=\Delta u-\Delta d$ at each $m_{f}$, shown in Figure~\ref{fig:mfvolgA}
for both lattice volumes and tabulated in
Tables~\ref{tab:results_dbw2_b087_8cub}-\ref{tab:results_dbw2_b087_16cub}.
$\Ga^{\rm lattice}$ evaluated on the
smaller volume is clearly smaller for each value of $m_{f}$, and the
difference increases as $m_f$ decreases.
In contrast, \(g_{_V}\) does not show much dependence on the volume.

To compare to our previous DWF results with the Wilson gauge action,
we plot the value of $(\Ga/\Gv)^{\rm lattice}$ as a function of
$(m_\pi/m_\rho)^2$ in Figure~\ref{fig:mfvolgAgV}.  The smaller volume
results using the DBW2 gauge action ($\beta=0.87$) are 
the same (within statistical errors) as
our previous results using the Wilson gauge action
($\beta=6.0$) on a slightly larger volume.  The large volume DBW2
results exhibit mild quark mass dependence while both smaller volume
results show a marked decrease toward the chiral limit.  We conclude
that our previous DWF-Wilson-gauge-action results were significantly adversely
affected by finite volume.

Finally, we extrapolate $\Ga^{\rm ren}$ to the chiral limit.
For this purpose, we have two methods.  One is to extrapolate the charge
ratios $(\Ga/\Gv)^{\rm lattice}$ to the chiral limit where the relation $\Zv =
\Za$ is valid. The second method is the conventional one utilized in all other
calculations~\cite{
{Fukugita:1995fh}, {Liu:1994ab}, {Gockeler:1996wg},{Gusken:1999as},{Dolgov:2002zm},
{Capitani:1999zd},{Horsley:2000pz}}.  The chiral
extrapolation is performed on $\Ga^{\rm lattice} \times \Za$. Recall
that the latter requires the value of $\Za$, whether nonperturbatively or
perturbatively calculated, while the former does not.  In the present case, we
use the nonperturbative value of $\Za$ from~\cite{Aoki:2002vt}.

We plot $(\Ga/\Gv)^{\rm lattice}$ and $\Za\times \Ga^{\rm lattice}$
together in Figure~\ref{fig:AlternategAgV} and perform a simple linear
extrapolation in each case.  The two methods provide consistent
results in the chiral limit: the ratio method gives $\Ga^{\rm
ren}=1.212 (27)$ while the conventional method gives $\Ga^{\rm
ren}=1.188 (25)$. In light of our earlier discussion, the systematic
difference if there is one, is related to our choice of
renormalization. A two percent error stemming from $\Zv \neq \Za$
yields 0.024. This is also the difference in the central values just
obtained. Thus, we quote
%
%
\beqn
\Ga^{\rm ren}=1.212 \pm 0.027 ({\rm stat}) \pm 0.024 ({\rm norm})
\eeqn
which underestimates the experimental value of 1.267 by less than five
percent.  We have not attempted to estimate residual non-zero lattice
spacing, finite volume, explicit chiral symmetry breaking, and
quenching effects. The first three are probably
small~\cite{Blum:2000kn,Aoki:2002vt,AliKhan:2000iv}. The only
remaining error not under good control is the quenching one which does not appear
to be large~\cite{{Gusken:1999as},{Dolgov:2002zm}}, in light of the relatively good agreement with
experiment shown above. This view does not change unless significant
non-analytic behavior, which we did not detect here, arises near the
chiral limit.  

We note Jaffe recently showed that in the chiral limit the nucleon axial charge is delocalized, and he argued this leads to a large reduction in $g_{_A}$ calculated in a finite volume surrounding the nucleon \cite{Jaffe:2001eb}.  Subsequently, Cohen showed that in a finite volume with 
periodic boundary conditions pertaining to lattice calculations, this phenomenon does not lead to a reduction in $g_A$ \cite{Cohen:2001bg}.  However, as emphasized in \cite{Cohen:2001bg}, this does not preclude other large finite volume effects.

As mentioned above, this calculation of $\Ga$ is performed for relatively
heavy quark masses; the quenching error at this unphysically large mass scale
is probably small.  However, one may worry that such a calculation does not
capture relevant physics in the region where the quark mass is much
lighter, and the so-called ``pion cloud'' surrounding the nucleon becomes
important.  Nevertheless the values of $\Ga^{\rm ren}$ at these
heavier quark masses already lie just a few percent below the experimental
value and show little dependence on the quark mass.  This presents an
important question concerning the role of the pion cloud: is it a few percent
effect, as seems plausible from our first principles calculation, or is it
larger, as estimated from phenomenological models~\cite{Detmold:2002nf}.


The dependence of the product \(m_{_N} g_{_A}\) on the lattice volume is of interest (See Fig.\ 
\ref{fig:mNgAren}).  While the smaller volume results always lie below the larger volume ones, within one standard deviation they almost always agree.  There is only one exception at \(a m_f=0.08\) in the bare lattice result.  No volume dependence is detected.  This is in clear contrast to the situation of the axial charge alone.  Since the product is the one that appears in the Goldberger-Treiman relation, it would be interesting to see how its counterpart, the induced pseudo-scalar form factor, behaves at small momentum transfer.

%% file: text_sections/outline/outline_III.tex
\ifnum\theOutline=1
\noindent \framebox{Begin \ outline \hspace{5.0in} }
\begin{enumerate}
\item Numerical calculations with Wilson (sigle plaquette) gauge action:
\begin{itemize}
\item RIKEN-BNL-Columbia QCDSP,
\item 400 gauge configurations, using a heat-bath algorithm,
\item \(\beta=6.0\), \(16^3\times 32 \times 16\), \(M_5 = 1.8\),
\item source at \(t=5\), sink at 21, current insertions in between.
\end{itemize}

\item \(Z_{_V} = 1/g_{_V}^{\rm lattice}\) is well-behaved,

\item \(g_{_A}/g_{_V}\): averaged in \(10\le t \le 16\),

\item While relevant three-point functions are well behaved in DWF, and
\(Z_{_V} = Z_{_A}\) is well satisfied, \(0.760(7)\) and
\(0.7555(3)\).
\item Why so small?
\begin{itemize}
\item finite lattice volume~\cite{{Jaffe:2001eb},{Cohen:2001bg}},
\item excited states (small separation between
\(t_{\rm source}\) and \(t_{\rm sink}\)),
\item quenching (zero modes, absent pion cloud, ...).
\end{itemize}

\item To investigate size-dependence, we need
\begin{itemize}
\item good chiral behavior, {\it i.e.\/} close enough to the
continuum, and
\item big enough volume.
\end{itemize}

\item Improved gauge actions help both.  DBW2\footnote{QCD-TARO
collaboration, Nucl.\ Phys.\ B577, 263 (2000); RBC collaboration, in preparation.}, in particular,
\[
S_G = \beta [ c_0 \sum W_{1,1} + c_1 \sum W_{1,2} ],
\]
with \(c_0 +8 c_1 = 1\) and \(c_1 = -1.4069\):
\item Bare \(g_{_A}^{\rm lattice}\) from wall source show volume dependence at medium \(m_f\) (\((2.4 {\rm fm})^3\) (filled) and \((1.2 {\rm fm})^3\) (open) volumes):

\item Bare \(g_{_V}^{\rm lattice}\) from sequential source (\((2.4 {\rm fm})^3\)):
\item \((g_{_A}/g_{_V})^{\rm lattice}\) from sequential source (\((2.4 {\rm fm})^3\)):
\item \((g_{_A}/g_{_V})^{\rm lattice} = (g_{_A}/g_{_V})^{\rm continuum}\): \(m_f\) and volume dependence in physical scale (set by \(m_\rho\)):
\end{enumerate}
\noindent \framebox{End \ outline \hspace{5.0in} }
\vspace{0.25in}
\fi

%% file: text_sections/outline/outline_IIIa.tex
\ifnum\theOutline=1
\noindent \framebox{Begin \ outline \hspace{5.0in} }
\begin{enumerate}
\item Numerical calculations with Wilson (sigle plaquette) gauge action:
\begin{itemize}
\item RIKEN-BNL-Columbia QCDSP,
\item 400 gauge configurations, using a heat-bath algorithm,
\item \(\beta=6.0\), \(16^3\times 32 \times 16\), \(M_5 = 1.8\),
\item source at \(t=5\), sink at 21, current insertions in between.
\end{itemize}

\item \(Z_{_V} = 1/g_{_V}^{\rm lattice}\) is well-behaved, see Fig.\ \ref{fig:ZV},
\begin{itemize}
\item the value \(0.764(2)\) at \(m_f = 0.02\) agrees well with
\(Z_{_A} = 0.7555(3)\) from
\begin{itemize}
\item \(\langle A^{\rm conserved}_\mu (t) \bar{q} \gamma_5 q (0) \rangle =
Z_{_A} \langle A^{\rm local}_\mu (t) \bar{q} \gamma_5 q (0) \rangle\)~\cite{Blum:2000kn},
\end{itemize}
\item linear fit gives \(Z_{_V} = 0.760(7)\) at \(m_f = 0\),
and quadratic fit, 0.761(5).
\end{itemize}
\item \(g_{_A}/g_{_V}\): averaged in \(10\le t \le 16\), Fig.\ \ref{fig:gA},
\begin{itemize}
\item linear extrapolation yields 0.81(11) at \(m_f=0\), and simlarly
small values for
\begin{itemize}
\item \(\Delta q /g_{_V} = 0.49(12)\) and
\item \((\delta q / g_{_V})^{\rm lattice} = 0.47(10)\) (with a preliminary
\(Z_{_T} \sim 1.1\)).
\end{itemize}
\end{itemize}

\end{enumerate}
\noindent \framebox{End \ outline \hspace{5.0in} }
\vspace{0.25in}

\fi

%% file: text_sections/outline/outline_IIIb.tex
\ifnum\theOutline=1
\noindent \framebox{Begin \ outline \hspace{5.0in} }
\begin{enumerate}
\item Improved gauge actions help both.  DBW2\footnote{QCD-TARO
collaboration, Nucl.\ Phys.\ B577, 263 (2000); RBC collaboration, in preparation.}, in particular,
\[
S_G = \beta [ c_0 \sum W_{1,1} + c_1 \sum W_{1,2} ],
\]
with \(c_0 +8 c_1 = 1\) and \(c_1 = -1.4069\):
\begin{itemize}
\item very small residual chiral symmetry breaking, \(am_{\rm res} < 10^{-3}\),
\item at the chiral limit, \(am_\rho = 0.592(9)\) (so \( a^{-1} \sim 1.3 {\rm GeV}\)), \(m_\rho/m_N \sim 0.8\), 
\item \(m_\pi (m_f=0.02) \sim 0.3 a^{-1}\).
\end{itemize}
\item DBW2 calculations are performed at \(a\) \(\sim\) 0.15 fm (\(\beta
= 0.87\)) with both wall and sequential sources on
\begin{itemize}
\item \(8^3 \times 24 \times 16\) (\(\sim\) \((1.2 {\rm fm})^3\)), 405 configurations (wall) and 405 (sequential),
\item \(16^3 \times 32 \times 16\) (\(\sim\) \((2.4 {\rm fm})^3\)), 100 configurations  (wall) and 416 (sequential),
\item source-sink separation of about 1.5 fm,
\item \(m_f\) = 0.02, 0.04, ...: \(m_\pi \ge 390 {\rm MeV}\), \(m_\pi L\ge 4.8\) and 2.4.
\end{itemize}
\item Renormalization factors: \({\cal O}^{\rm continuum}(\mu) = Z_{\cal
O}(a\mu) {\cal O} ^{\rm lattice} (a)\)., Fig.\ \ref{fig:ZAV}:
\begin{itemize}
\item \(Z_{_V}\) shows slight quadratic dependence on \(m_f\) as
expected: \(V_\mu^{\rm conserved} = Z_{_V} V_\mu^{\rm local} + {\cal O}
(m_f^2 a^2)\),
\begin{itemize}
\item yielding a value \(Z_{_V} = 0.7951(11)\) from linear fit, ( 0.7990(28) from quadratic fit)
\item agrees well with \(Z_{_A} = 0.77759(45)\) \footnote{RBC Collaboration, in preparation: this value is obtained from a relation
\(\langle A_{_\mu}^{\rm conserved} (t) [\bar{q}\gamma_5 q](0) \rangle =
Z_{_A} \langle A_\mu^{\rm local} (t) [\bar{q}\gamma_5 q](0) \rangle\).} if  the expected \( {\cal O}(a^2) \) error
is taken into account, See Fig.\ \ref{fig:ZVSYS}.
\end{itemize}
\end{itemize}
\item Bare \(g_{_A}^{\rm lattice}\) from wall source show volume dependence at medium \(m_f\) (\((2.4 {\rm fm})^3\) (filled) and \((1.2 {\rm fm})^3\) (open) volumes), Fig.\ \ref{fig:BaregA}.
\item Bare \(\Delta u^{\rm lattice}\) and \(\Delta d^{\rm lattice}\) from sequential source (\((2.4 {\rm fm})^3\)), FIg.\ \ref{fig:BareDeltaq}:

\item Bare \(g_{_V}^{\rm lattice}\) from sequential source (\((2.4 {\rm fm})^3\)), Fig.\ \ref{fig:BaregV}.
\item \((g_{_A}/g_{_V})^{\rm lattice}\) from sequential source (\((2.4 {\rm fm})^3\)): Fig.\ \ref{fig:SequentialgAgV}.
\item \((g_{_A}/g_{_V})^{\rm lattice} = (g_{_A}/g_{_V})^{\rm continuum}\): \(m_f\) and volume dependence in physical scale (set by \(m_\rho\)), Fig.\ \ref{fig:mfvolgAgV}:
\begin{itemize}
\item Clear volume dependence is seen between \((2.4 {\rm fm})^3\) and \((1.2 {\rm fm})^3\) volumes.
\item The large volume results (sequential)
\begin{itemize}
\item show a very mild \(m_f\) dependence,
\item extrapolate to about 5 \% under estimation, \(g_{_A} = 1.21(3)\).
\end{itemize}
\item \(1/({\rm volume})\) extrapolation to infinite volume suggests +0.04 systematics.
\end{itemize}
\item Alternatively we can use \(g_{_A}^{\rm lattice} \times Z_{_A}\), Fig.\ \ref{fig:AlternategAgV}:
agree well with \((g_{_A}/g_{_V})^{\rm lattice}\) in the chiral limit, and and an expected difference seen away from there.  This leads to a systematic error estimate of \(\pm 0.03\), which stems from the ${\cal O}(a^2)$ in determination of
the renormalization factor $Z_{_V} (Z_{_A})$.
\end{enumerate}
\noindent \framebox{End \ outline \hspace{5.0in} }
\vspace{0.25in}
\fi

%% file: text_sections/conclusions.tex
%
%
\input{text_sections/outline/outline_IV.tex}

In this paper we have studied 
the nucleon axial charge 
and the
vector charge in quenched lattice QCD. To capture important aspects of the
chiral symmetry of QCD, we used domain wall fermions to simulate the light
quarks.

We first demonstrated that the lattice renormalization of the isovector vector
and axial-vector currents satisfy \(\Zv=\Za\) to a high degree of precision,
less than a percent at $a^{-1}\approx 2$ GeV and about two percent at
$a^{-1}\approx 1.3$ GeV.  This is achieved because in practice the DWF method
preserves the chiral symmetry of QCD up to small corrections and hence
maintains the relevant Ward-Takahashi identity. This holds if the underlying
(quenched) gauge configuration is sufficiently smooth.  Ensembles of such
gauge configurations are obtained 
close to the continuum limit.
For the single-plaquette Wilson gauge action $\beta=6.0$ 
($a^{-1}\approx 2$ GeV) is good enough.  For the DBW2 action the lattice
spacing may be significantly larger while still maintaining good chiral
symmetry ($a^{-1}\approx 1.3$ GeV).

Our first calculation of $\Ga$ with the Wilson gauge action was performed at
$\beta =6.0$ on a $16^3 \times 32$ lattice.  The corresponding spatial volume
$\sim (1.6 {\rm fm})^3$ is similar to those used in previous lattice
calculations.  This volume is rather small in comparison with the
experimentally measured proton charge radius.  On this lattice we found all
the relevant three-point functions 
are well behaved and that we 
can reliably extract the charges.  The isovector vector
current renormalization, $\Zv$, determined from them agrees well with the
corresponding axial current renormalization, $\Za$, independently obtained
from the axial Ward-Takahashi identity.  We found that both the axial charge,
$\Ga^{\rm lattice}$, and its ratio to the vector charge, $(\Ga/\Gv)^{\rm
lattice}$, exhibit a very strong dependence on the quark mass.  A simple
linear extrapolation of $(\Ga/\Gv)^{\rm lattice}$ to zero quark mass
yielded a very small value, about 2/3 of the experimental one.

The second quenched calculation employed the DBW2 gauge action with 
$\beta=0.87$ set for a coarser lattice spacing, $a\approx 0.15$ fm.  This
allowed a larger physical volume while maintaining good chiral symmetry. To
study pure finite volume effects, at fixed lattice spacing we calculated on
lattices with sizes $16^3 \times 32$ and $8^3 \times 24$ ($\sim (1.2 {\rm
fm})^3$ and $\sim (2.4 {\rm fm})^3$, respectively).  A significant dependence
on the volume is seen in both the axial charge $\Ga^{\rm lattice}$ and the
charge ratio $(\Ga/\Gv)^{\rm lattice}$, with the larger volume giving larger
values.  %
In contrast \(g_{_V}\) does not show such dependence.
The dependence on the quark mass is also different.  In the larger
volume the central values remain almost constant, while in the smaller volume
they decrease noticeably with the quark mass.  In the chiral limit the two
differ by about 20\% difference.  The behavior of $\Ga$ on the smaller volume
is quite consistent with that observed in the earlier calculation with the
Wilson gauge action.

Our estimate of $(\Ga/\Gv)^{\rm ren}$ at zero quark mass from the
larger volume with DBW2 action is $1.212 \pm 0.027 ({\rm stat}) \pm 0.024
({\rm norm})$.  The systematic error is estimated from the two percent
difference between $\Za$ and $\Zv$ which are theoretically equivalent to
${\cal O}(a^2)$, neglecting even smaller effects induced by explicit
chiral symmetry breaking in DWF.   It underestimates the experimental value of
1.2670(30) by less than five percent.  This discrepancy is smaller than twice
the theoretical error. 

Thus dependence on the volume seems to be the largest among the known sources
of systematic error for the first principles lattice calculation of $\Ga$.
This suggests
that close attention be paid to the finite volume effect in other lattice
numerical studies of nucleon structure, in particular the moments of
spin-polarized structure functions which are related to the axial charge.

It should be also noted that although the crucial relation $\Zv=\Za$ is
satisfied well in the second calculation, we detected 
small ${\cal O}(a^2)$
differences among different determinations of $\Zv$.  Such differences were
not detectable in the first set of simulations with the Wilson gauge action at
$a\approx 0.10$ fm.  Numerically this is at most a few percent effect, and
does not affect the volume dependence.

As discussed at the end of section III, the present calculation of $\Ga$ was
performed using relatively heavy quark masses (390 MeV $\le m_{\pi} \le$ 860
MeV) so that the systematic error arising from quenching may be small.
However, one may worry that such a calculation does not capture the physics of
the pion cloud surrounding the nucleon.  In spite of this, the values of
$\Ga^{\rm ren}$ at these unphysically heavy quark masses lie just below the
experimental value and show little, if any, dependence on the quark mass.
It is also interesting to note that the product $m_{_N} \Ga$ shows
noticeably less volume dependence.  Since this is the combination that appears
in the Goldberger-Treiman relation, it would be interesting to see how its
counterpart, the induced pseudo-scalar form factor, behaves at small but
non-zero nucleon momentum.  

%% file: text_sections/outline/outline_IV.tex
\ifnum\theOutline=1
\noindent \framebox{Begin \ outline \hspace{5.0in} }
\begin{enumerate}
\item Conclusions: with quenched DBW2 and DWF for nucleon currents, indications
are seen for
\begin{itemize}
\item good chiral behavior:
\begin{itemize}
\item especially the relation \(Z_{_A} = Z_{_V}\) is easily and well
maintained
\item found ${\cal O}(a^2)$ error in determination of  \(Z_{_V}\) 
\end{itemize}
\item milder \(m_f\) dependence,
\item clear size dependence, a 20\% increase from 1.2 fm to 2.4 fm,
\begin{itemize}
\item with the product \(m_{_N}g_{_A}\) almost constant (or increase by 10 \%?),
\item while the \(m_{_\pi}/m_{_\rho}\) does not differ between the two volumes,
\end{itemize}
\item \(g_{_A}/g_{_V} = 1.21(3)_{\rm stat}(\pm3)_{\rm syst}\),
\begin{itemize}
\item this systematic error of \(\pm 0.03\) is from the \(Z_{_A}\) systematics alone,
\item and not the \(+0.04\) from \(1/V\) ansatz.
\end{itemize}
\end{itemize}
\item Future:
\begin{itemize}
\item a few more observables, e.g.\ interesting to see how well quenched calculation works for
a well-known example of soft-pion,
\begin{itemize}
\item Goldberger-Treiman relation: \(g_{_A}/g_{_V} \simeq f_\pi g_{\pi N}
/m_N\),
\end{itemize}
\item detailed study of volume dependence, like 1.8 fm,
\item flavor structure.
\end{itemize}
\end{enumerate}
\noindent \framebox{End \ outline \hspace{5.0in} }
\fi

%% file: text_sections/appendix.tex
%
%


\section{Current algebra and CVC hypothesis}
\label{sec:appendixA}

Defining the charge 
${\cal Q}_{_{V}}^{a}=i\int d^3 x V_{0}^a(\vec{x}, t)$
\protect\footnote{
A definition of ${\cal Q}_{_{V}}^{a}$ is slightly different from 
the one in \cite{Sasaki:2001nf}. Only difference is a factor of $\frac{i}{2}$.
}, 
the transformation for the quark fields in the isospin $SU(2)$
subgroup of the $SU(2)\times SU(2)$
chiral symmetry can be represented by
%
%
\barr
\left[{\cal Q}_{_{V}}^{a}, \psi(x)\right]&=&-T^{a}\psi(x), \\
\left[{\cal Q}_{_{V}}^{a}, {\bar \psi}(x)\right]&=&+{\bar \psi}(x)T^{a},
\earr
where $\psi=(u, d)^{T}$ and $V_{\mu}^a(x)={\bar \psi}(x)\gamma_{\mu}T^{a}\psi(x)$.  One
can easily find the axial current $A_{\mu}^a(x)={\bar
\psi}(x)\gamma_{5}\gamma_{\mu}T^{a}\psi(x)$ and vector current $V_{\mu}^a$
transform under isospin symmetry as
%
%
\barr
\left[{\cal Q}_{_{V}}^{a}, V_{\mu}^{b}(x)\right]&=&
i\varepsilon_{abc}V_{\mu}^{c}(x) \\
\left[{\cal Q}_{_{V}}^{a}, A_{\mu}^{b}(x)\right]&=&
i\varepsilon_{abc}A_{\mu}^{c}(x).
\earr
According to %
the above $SU(2)$ current algebra, $A_{\mu}^{+}$ and $V_{\mu}^{+}$ can be expressed as 
%
%
\barr 
A_{\mu}^{\pm}(x)&=&
{\bar \psi}(x)\gamma_{\mu}\gamma_{5}T^{\pm}\psi(x)=
-[{\cal Q}_{_{V}}^{\pm},A_{\mu}^{3}(x)],\\ 
V_{\mu}^{\pm}(x)&=&{\bar \psi}(x)\gamma_{\mu}T^{\pm}\psi(x)=
-[{\cal Q}_{_{V}}^{\pm},V_{\mu}^{3}(x)], \earr
where $T^{\pm}=T_{1}\pm iT_{2}$. Hence, under the CVC hypothesis, one can find
%
%
\barr \lproton A_{\mu}^{+}\rneutron &=& -\lproton [
{\cal Q}_{_{V}}^{+},A_{\mu}^{3}] \rneutron \nonumber \\ &=&-\lproton 
{\cal Q}_{_{V}}^{+}A_{\mu}^{3}\rneutron +\lproton A_{\mu}^{3}
{\cal Q}_{_{V}}^{+}\rneutron\nonumber \\ 
&=&\lproton A_{\mu}^{3}\rproton - \lneutron
A_{\mu}^{3}\rneutron \nonumber  \\ 
&=&2\lproton A_{\mu}^{3}\rproton.
\earr
The third line follows from
${\cal Q}_{_{V}}^{+}\rneutron =\rproton$ and $\lproton {\cal Q}_{_{V}}^{+}=\lneutron$.
A similar calculation for the vector case yields the relation, 
$\lproton V_{\mu}^{+}\rneutron =2\lproton V_{\mu}^{3}\rproton$.

\pagebreak

%% file: tab/tab.tex


\begin{table}[htbp]
\caption{Previous lattice calculations with Wilson or ${\cal O}(a)$ improved
Wilson (clover) fermions. In general, $\Ga$ is significantly underestimated.
Note that almost all volumes are small, most estimates use perturbative
renormalization, and partially-unquenching did not increase the 
value of $\Ga$. 
The \(\dagger\) symbol denote the
continuum extrapolated value.
}
\label{tab:ga_lattice_hist}
\begin{ruledtabular}
\begin{tabular}{llllllrllc}
\hline type &group&fermion &$L^{3}\times N_{t}$ &\(\beta\)
&volume&statistics&\(m_\pi L\)&\(g_{_A}\) & Ref. \\ \hline\hline quenched &KEK
&Wilson &\(16^3\times 20\)&5.7&\(({\rm 2.2 fm})^3\)&260&\(\ge 5.9\)&0.985(25)
&\cite{Fukugita:1995fh} \\

&Kentucky
&Wilson   &\(16^3\times 24\)&6.0&\(({\rm 1.5 fm})^3\)&24  &\(\ge 5.8\)&1.20(10)
&\cite{Liu:1994ab}\\
&DESY &Wilson &\(16^3\times 32\)&6.0&\(({\rm 1.5 fm})^3\)&1000&\(\ge
4.8\)&1.074(90) &\cite{Gockeler:1996wg} \\

&LHPC-SESAM &Wilson&\(16^3\times 32\)&6.0&\(({\rm 1.5 fm})^3\)&200&\(\ge
4.8\)&1.129(98) &\cite{Dolgov:2002zm}\\
&QCDSF &Wilson &\(16^3\times 32\)&6.0&\(({\rm 1.5
fm})^3\)&O(500)&&1.14(3)$^{\dagger}$ &\cite{Capitani:1999zd}\\ &&&\(24^3\times
48\)&6.2&\(({\rm 1.6 fm})^3\)&O(300)&&&\\ &&&\(32^3\times 48\)&6.4&\(({\rm 1.6
fm})^3\)&O(100)&&&\\ &QCDSF-UKQCD &Clover&\(16^3\times 32\)&6.0&\(({\rm 1.5
fm})^3\)&O(500)&&1.135(34)$^{\dagger}$ &\cite{Horsley:2000pz}\\
&&&\(24^3\times 48\)&6.2&\(({\rm 1.6 fm})^3\)&O(300)&&&\\ &&&\(32^3\times
48\)&6.4&\(({\rm 1.6 fm})^3\)&O(100)&&&\\ full(\(N_f=2\)) &LHPC-SESAM
&Wilson&\(16^3\times 32\)&5.5&\(({\rm 1.7 fm})^3\)&100&\(\ge 4.2\)&0.914(106)
&\cite{Dolgov:2002zm}\\ &SESAM &Wilson &\(16^3\times 32\)&5.6&\(({\rm 1.5
fm})^3\)&200 &\(\ge 4.5\)&0.907(20) &\cite{Gusken:1999as}\\ \hline
\end{tabular}
\end{ruledtabular}
\end{table}

\input{tab/tab1.tex}

\input{tab/tab2.tex}

\pagebreak

%% file: tab/tab1.tex
\begin{table}[htbp]
\label{tab:simulation_parameters}
\caption{
Simulation parameters for each action and each volume studied in this work.}
\label{tab:simulation_pamam}
\begin{ruledtabular}
\begin{tabular}{lccccc}
\hline
Gauge action & $\beta$ & $L^3 \times N_t$ & $L_s$ & $M_5$ & volume\\
\hline
Wilson    &  6.0  & $16^3 \times 32$ & 16 & 1.8 & $(1.6 {\rm \,fm})^3$\\ 
\hline
DBW2    & 0.87 & $16^3 \times 32$ & 16 & 1.8 &$(2.4 {\rm \,fm})^3$\\ 
    &  & $8^3 \times 24$ & 16 & 1.8 & $(1.2 {\rm \,fm})^3$\\ 
\hline
\end{tabular}
\end{ruledtabular}
\end{table}

\begin{table}[htbp]
\caption{More parameters for each action and each volume.}
\label{tab:simulation_information}
\begin{ruledtabular}
\begin{tabular}{lccll}
\hline Gauge action ($\beta$) & $L^3 \times N_t$ & quark mass values & 
statistics(type) & $m_{\pi}L$ \\ 
\hline Wilson (6.0) & $16^3 \times 32$ & 0.02, 0.03,
0.04, 0.05 & 400 (Wall) &$\ge$4.3\\ 
\hline DBW2 (0.87) & $16^3 \times 32$ & 0.02, 0.04, 0.06, 0.08, 0.10 & 416 (Sequential) & $\ge$4.8\\
 & $8^3 \times
24$ & 0.04, 0.06, 0.08, 0.10 & 400 (Sequential) & $\ge$3.4\\ 
& & 0.04, 0.06,
0.08, 0.10 & 400 (Wall) &\\ 
\hline
\end{tabular}
\end{ruledtabular}
\end{table}


\begin{table}[htbp]
\caption{The residual mass $m_{\rm res}$, hadron masses, inversre lattice spacing
($a^{-1}_{\rho}$, set by the $\rho$ meson mass) and the renormalization factor
of the axial-vector current ($\Za$).  The $\rho$ meson mass and the nucleon mass are
given in the chiral limit in each case.  }
\label{tab:fermionic_scale_param}
\begin{ruledtabular}
\begin{tabular}{lccccccccc}
Gauge action ($\beta$) & $M_5$ & $L_s$ & $m_{\rm res}$ & $m_{\rho}$ & $m_{N}$ & $a^{-1}_{\rho} 
$ (GeV) & $\Za$ & Ref.\\ 
\hline
Wilson (6.0) & 1.8 & 16 & 1.24 (5) $\times 10^{-3}$ &  0.404 (8) & 0.566 (21) &
1.922 (40) & 0.7555 (3) &
\cite{Blum:2000kn} \\
DBW2 (0.87) & 1.8 & 16 & 5.69 (26) $\times 10^{-4}$ &  0.589 (19) & 0.780 (27)
& 1.31 (4) & 0.77759 (45) &
\cite{Aoki:2002vt} \\
\hline
\end{tabular}
\end{ruledtabular}
\end{table}



\begin{table}[htbp]
\caption{Hadron masses computed using Wilson gauge action
at $\beta = 6.0$, $16^3 \times 32$, $M_5=1.8$, $L_s=16$, from Ref.\  \cite{Sasaki:2001nf}.} 
\label{tab:hadrons_wilson_b60}
\begin{ruledtabular}
\begin{tabular}{llll}
\hline
$m_f$ & $m_{\pi}$ &  $m_{\rho}$ & $m_{N}$ \\
\hline\hline
0.02&    0.2687 (24)   & 0.4530 (62) & 0.645 (12) \\
0.03&    0.3224 (21)   & 0.4814 (45) & 0.716 (5)  \\
0.04&    0.3691 (19)   & 0.5126 (42) & 0.754 (6) \\
0.05&    0.4116 (18)   & 0.5395 (36) & 0.805 (5) \\
\hline
\end{tabular}
\end{ruledtabular}
\end{table}


\begin{table}[htbp]
\caption{Results for the nucleon axial charge, $\Delta u$, $\Delta d$
and $Z_{_V}=1/\Gv^{\rm latt}$, Wilson gauge action,
$\beta = 6.0$, $16^3 \times 32$, $M_5=1.8$, $L_s=16$, 400 configurations.
} 
\label{tab:results_wilson_b60}
\begin{ruledtabular}
\begin{tabular}{llllll}
\hline
$m_f$ & $(g_{_A})_{\rm latt}$ &  $(g_{_A})_{\rm ren}$
& $(\Delta u)_{\rm ren}$ & $(\Delta d)_{\rm ren}$ & $Z_{_V}$ \\
\hline\hline
0.02&    1.216 (106)   &0.929 (82)      &0.739 (82)      &-0.189 (40)& 0.7637 (23)\\
0.03&    1.380 (76)      &1.056 (59)      &0.840 (54)      &-0.215 (26)& 0.7654 (17)\\       
0.04&    1.480 (59)      &1.135 (46)      &0.903 (40)      &-0.232 (19)& 0.7671 (13)\\
0.05&    1.542 (47)      &1.186 (37)      &0.942 (32)      &-0.243 (15)& 0.7689 (11)\\
\hline
\end{tabular}
\end{ruledtabular}
\end{table}


%% file: tab/tab2.tex

\begin{table}[htbp]
\caption{Hadron masses computed using the DBW2 gauge action.
All fits for $\rho$ meson have $\chi^2 /N_{DF}<1.0$, and $<1.5$ for pion and nucleon.
}
\label{tab:masses_dbw2_b087}
\begin{ruledtabular}
\begin{tabular}{lllll}
\hline
$L^3 \times N_{t}$ & $m_f$ & $m_{\pi}$ & $m_{\rho}$ & $m_{N}$ \\
\hline\hline
 &0.04 &0.4255 (38) & 0.679  (16) & 1.071 (13) \\
$8^3 \times 24$ 
 &0.06 &0.5094 (34) & 0.729  (9) & 1.127 (11) \\
 &0.08 &0.5865 (30) & 0.776  (7) & 1.205 (9) \\
 &0.10 &0.6567 (27) & 0.823  (6) & 1.292 (10) \\
\hline
 & 0.02 & 0.3015  (16) & 0.647 (22) & 0.854  (6) \\
 & 0.04 & 0.4146  (16) & 0.681 (10)  & 0.963  (5) \\
$16^3 \times 32$
 & 0.06 & 0.5050  (16) & 0.725 (6) & 1.060 (4) \\
 & 0.08 & 0.5834  (15) & 0.771 (5) & 1.156 (4) \\
 & 0.10 & 0.6546  (14) & 0.819 (4) & 1.242 (4) \\
\end{tabular}
\end{ruledtabular}
\end{table}

\begin{table}[htbp]
\caption{Results for the nucleon axial charge, $\Delta u$, $\Delta d$
and $Z_{_V}=1/g_{_V}^{\rm latt}$, DBW2 gauge action, 
$\beta = 0.87$, $8^3 \times 24$, $M_5=1.8$, $L_s=16$, 400 configurations.
} 
\label{tab:results_dbw2_b087_8cub}
\begin{ruledtabular}
\begin{tabular}{llllll}
\hline
$m_f$ & $(g_{_A})_{\rm latt}$ &  $(g_{_A})_{\rm ren}$
& $(\Delta u)_{\rm ren}$ & $(\Delta d)_{\rm ren}$ & $Z_{_V}$ \\
\hline\hline
0.04&    1.303 (146)    &1.059 (120)    &0.690 (99)     &-0.369 (99)   &0.8191(65)\\
0.06&    1.342 (74)      &1.099 (62)       &0.817 (52)     &-0.282 (45)   &0.8242 (35)\\
0.08&    1.373 (46)      &1.136 (39)       &0.876 (32)     &-0.260 (25)   &0.8317 (24) \\
0.10&    1.398 (30)      &1.165 (26)       &0.902 (21)     &-0.263 (15)   &0.8403 (18)\\
\end{tabular}
\end{ruledtabular}
\end{table}

\begin{table}[htbp]
\caption{Results for the nucleon axial charge, $\Delta u$, $\Delta d$
and $Z_{_V}=1/g_{_V}^{\rm latt}$, DBW2 gauge action, 
$\beta = 0.87$, $16^3 \times 32$, $M_5=1.8$, $L_s=16$, 416 configurations.
} 
\label{tab:results_dbw2_b087_16cub}
\begin{ruledtabular}
\begin{tabular}{llllll}
\hline
$m_f$ & $(g_{_A})_{\rm latt}$ &  $(g_{_A})_{\rm ren}$
& $(\Delta u)_{\rm ren}$ & $(\Delta d)_{\rm ren}$ & $Z_{_V}$ \\
\hline\hline
0.02&    1.531 (60)      &1.229 (49)      &0.945 (44)      &-0.284 (27)   &0.8040 (19)\\
0.04&    1.523 (24)      &1.230 (20)      &0.946 (17)      &-0.284 (10)   &0.8115 (9)\\
0.06&    1.510 (15)      &1.230 (12)      &0.953 (10)      &-0.277 (6)     &0.8184 (6)\\
0.08&    1.505 (11)      &1.236 (9)        &0.963 (7)       &-0.273 (4)       &0.8260 (5) \\
0.10&    1.503 (8)        &1.246 (7)        &0.975 (6)       &-0.271 (3)       &0.8347 (5)\\
\end{tabular}
\end{ruledtabular}
\end{table}

%% file: fig/fig.tex
\begin{figure}
\begin{center}\begin{minipage}[b]{5.25in}
\includegraphics[width=2.5in]{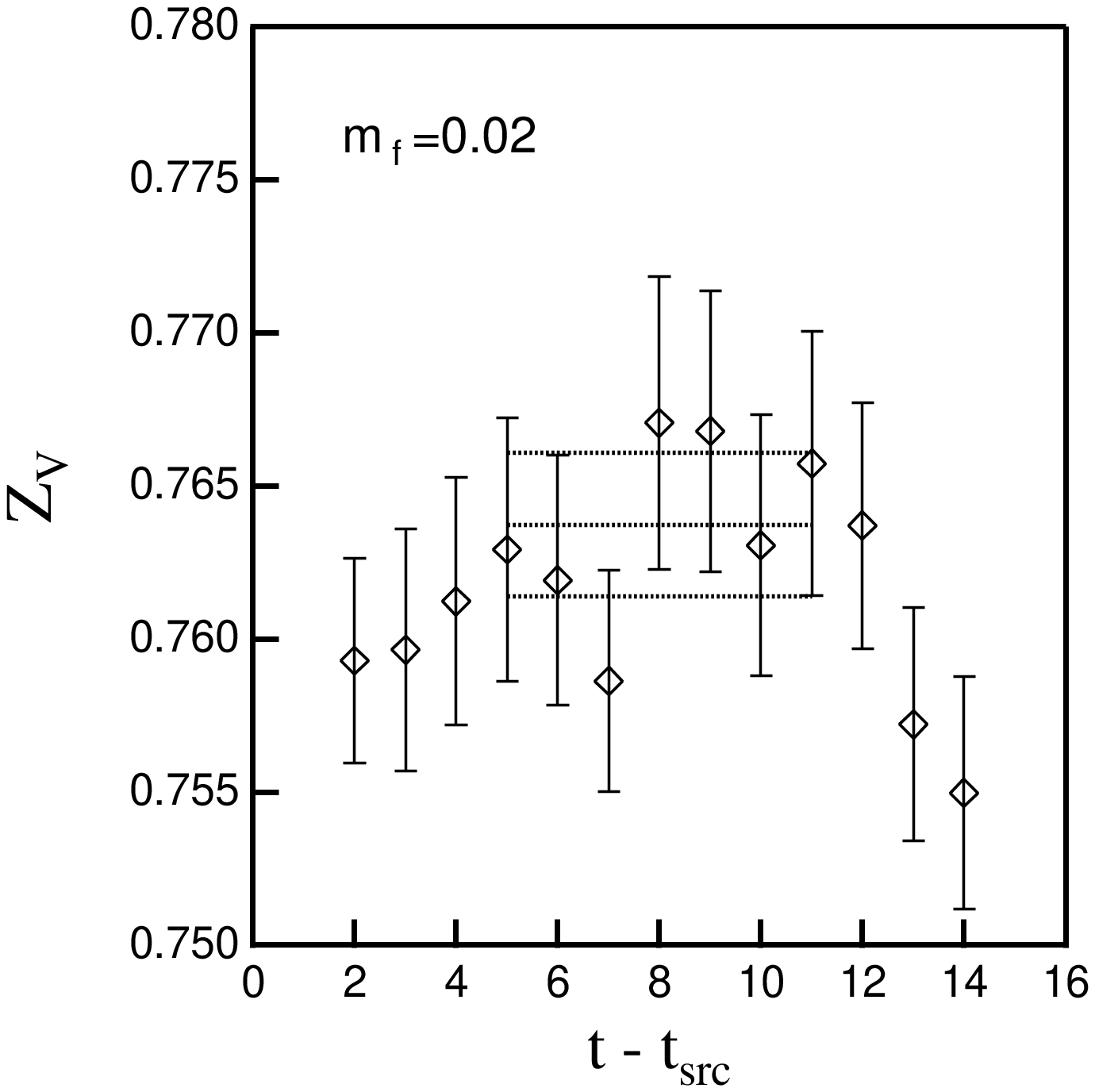}
\includegraphics[width=2.5in]{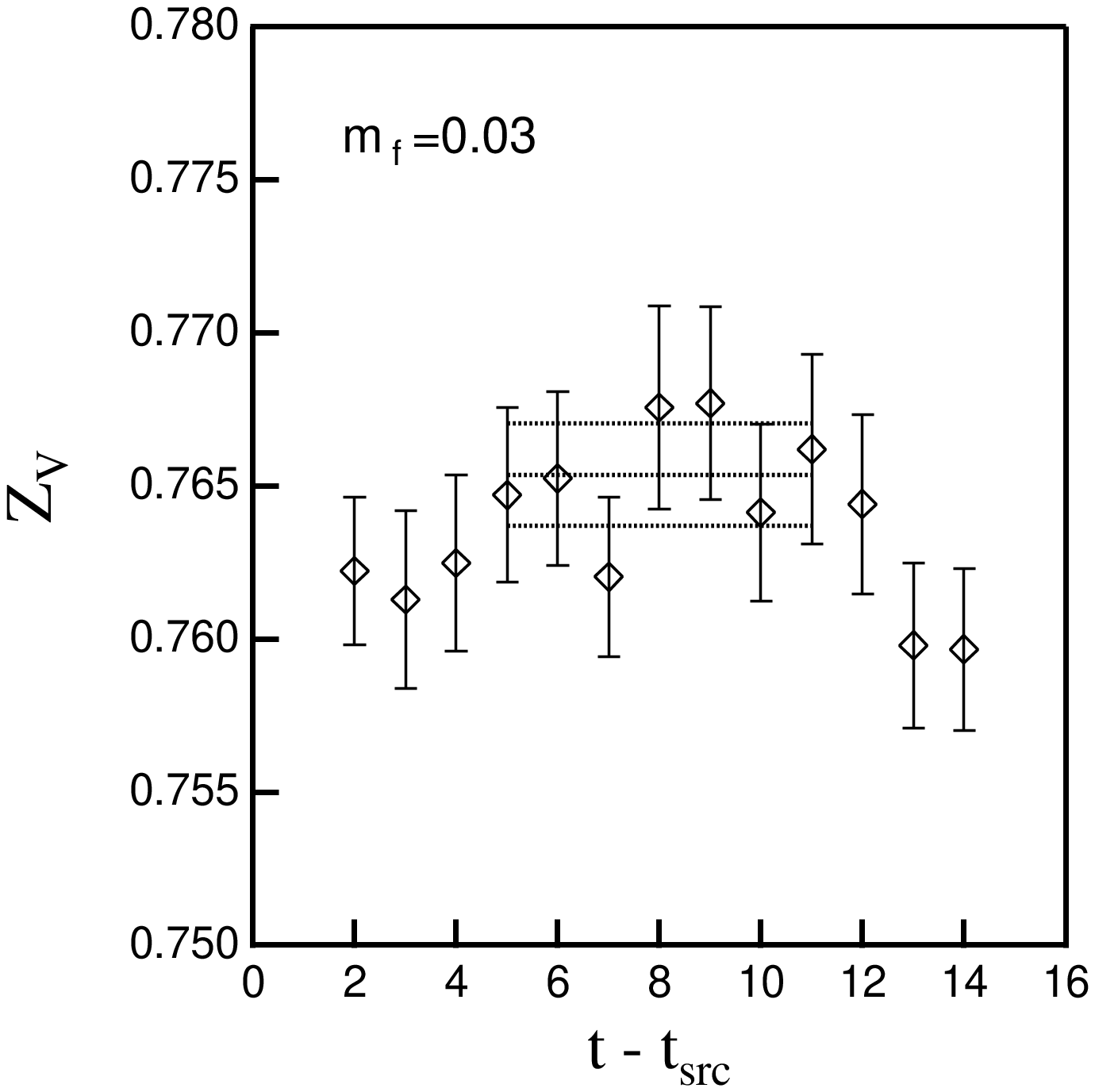}
\linebreak\vfill\includegraphics[width=2.5in]{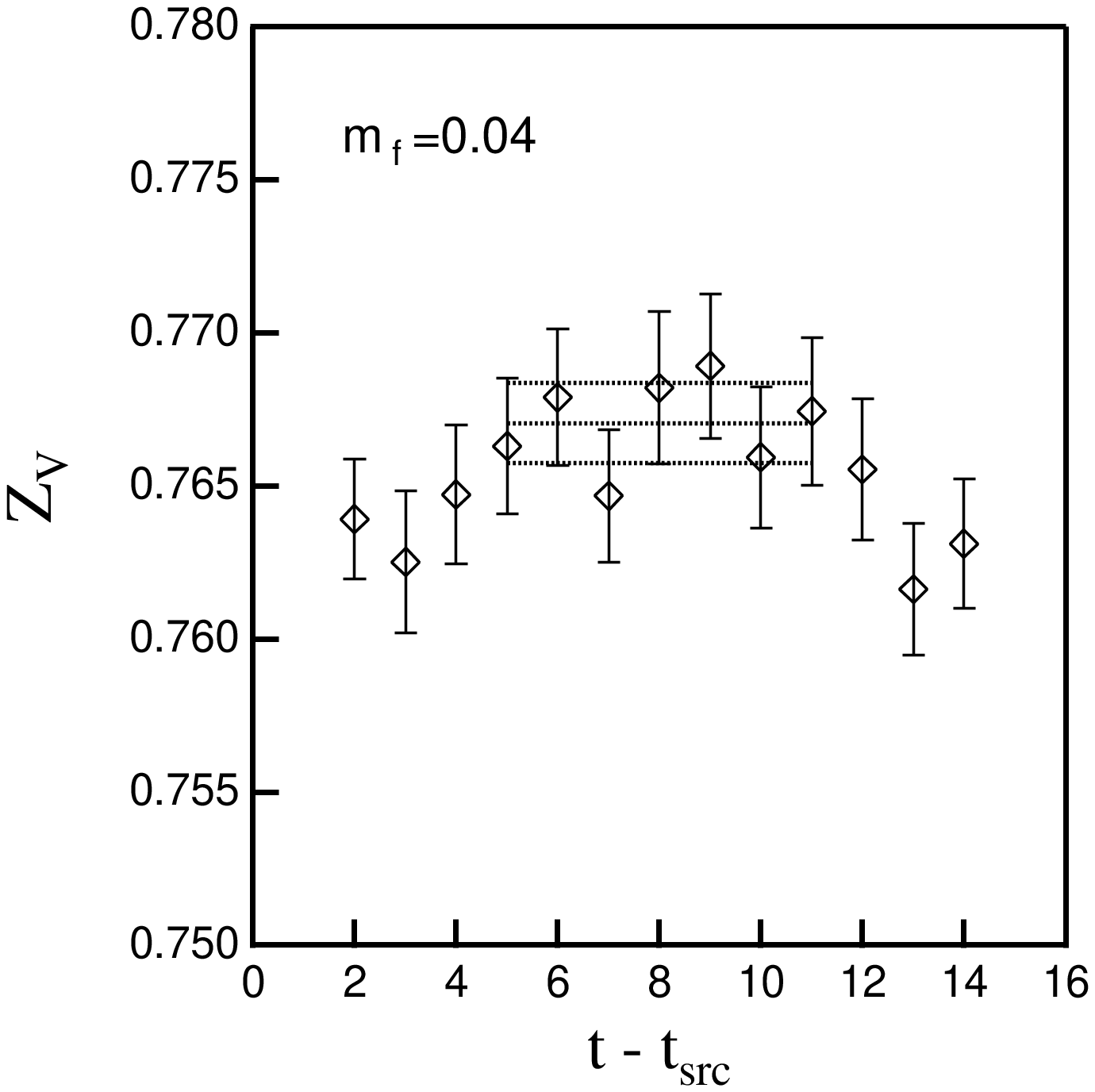}
\includegraphics[width=2.5in]{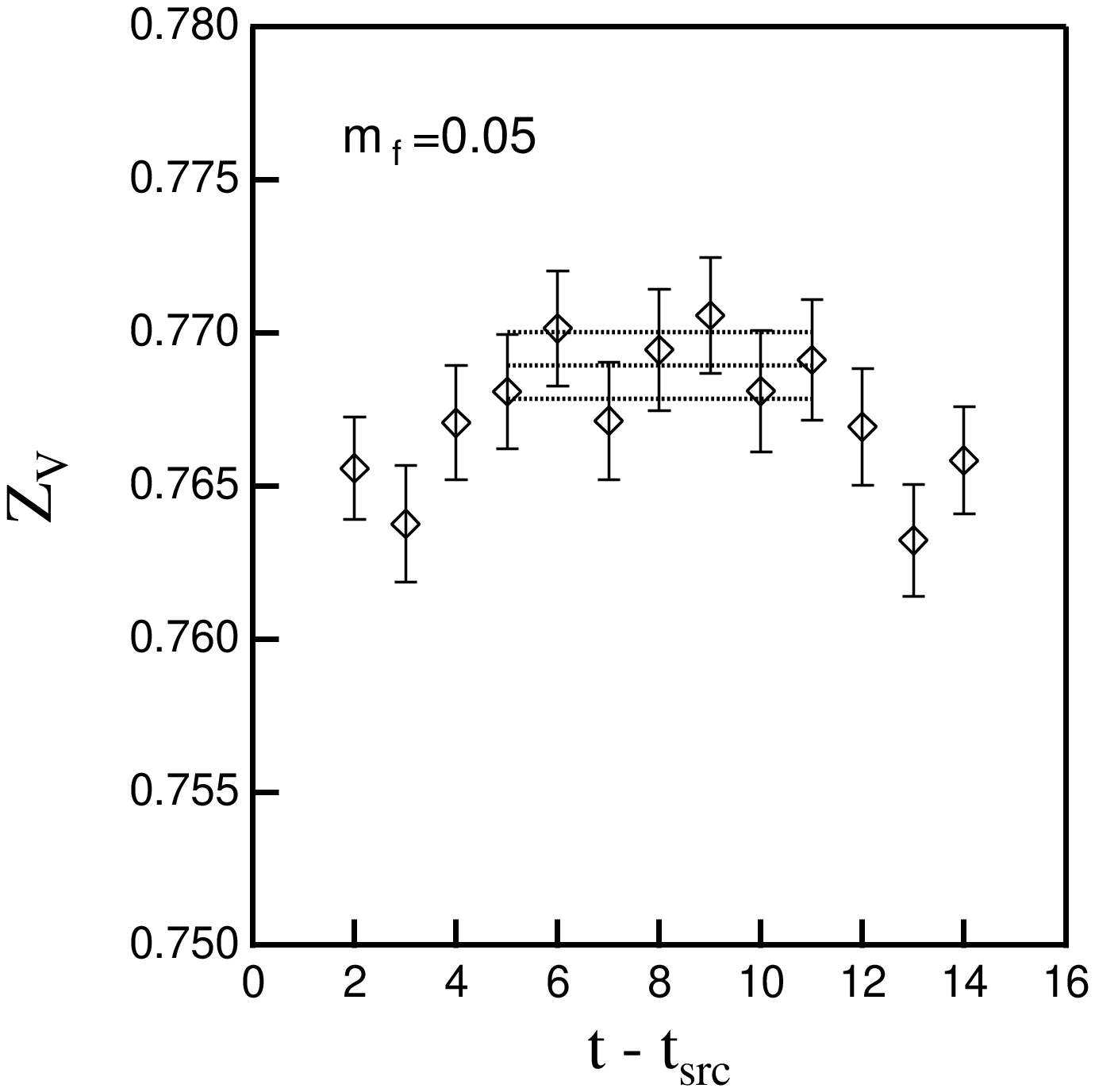}
\linebreak\vfill
\end{minipage}
\end{center}
\caption{$\Zv=1/\Gv^{\rm latt}$, Wilson gauge action, as a function of the current
insertion time-slice.  
Note the very fine scale.
A good plateau for each quark mass is observed in the
middle range between the source and sink. 
The lines denote central values and statistical errors from 
constant fits over the plateaus.}
\label{fig:ZvInsertion_W6}
\end{figure}

\begin{figure}\begin{center}
\includegraphics[height=5in]{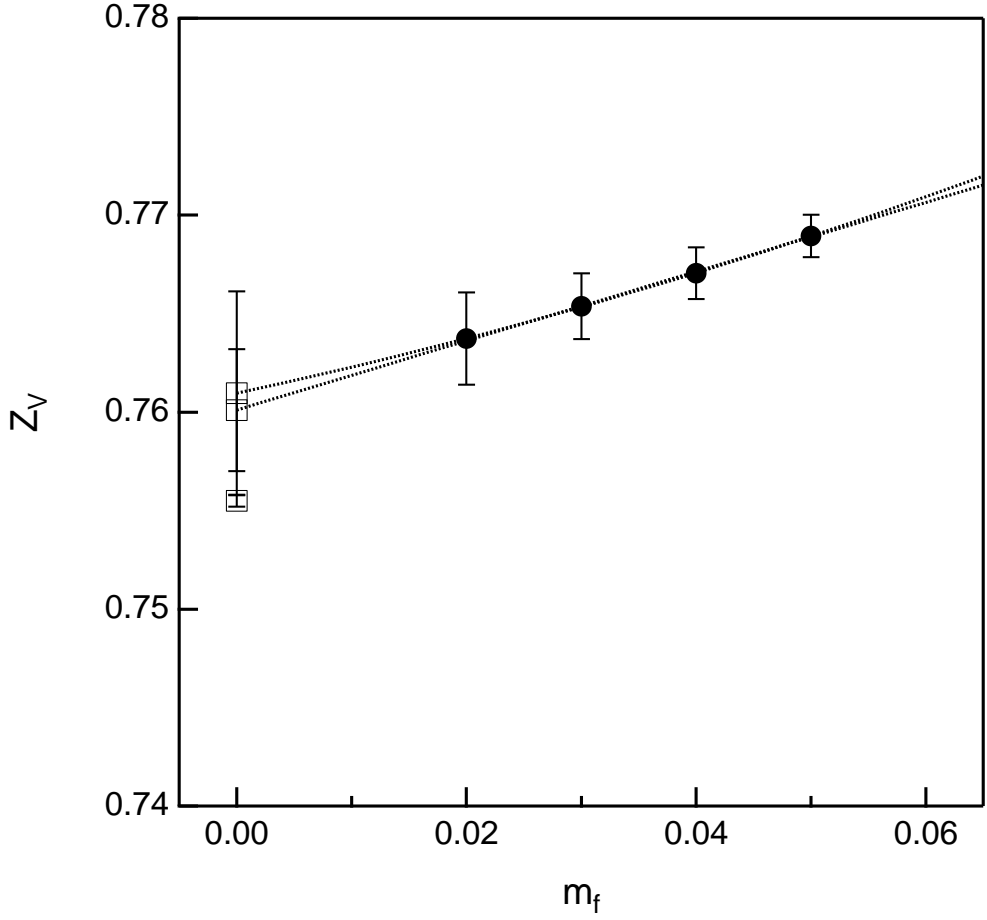}\end{center}
\caption{Quark mass dependence of the vector current renormalization, $\Zv=
1/\Gv^{\rm latt}$,
Wilson gauge action (note scale). Lines denote uncorrelated linear and
quadratic fits to the data points. Extrapolated values are consistent with the
axial-vector current renormalization computed from meson two-point
functions~\cite{Blum:2000kn,Blum:2001sr} to less than a percent.}
\label{fig:Zv_mass_dep_W6}
\end{figure}

\begin{figure}
\begin{center}\begin{minipage}[b]{5.25in}
\includegraphics[width=2.5in]{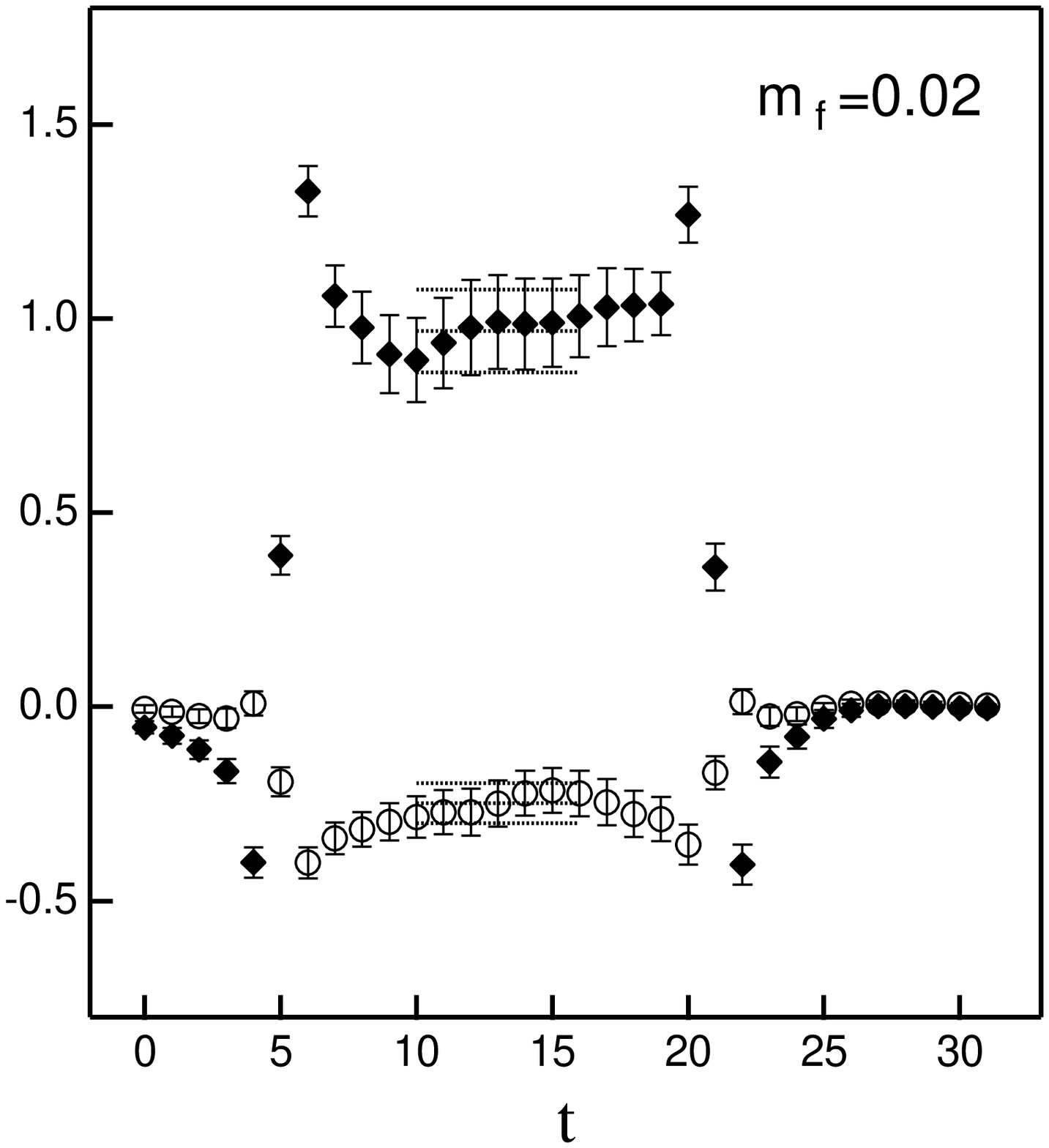}
\includegraphics[width=2.5in]{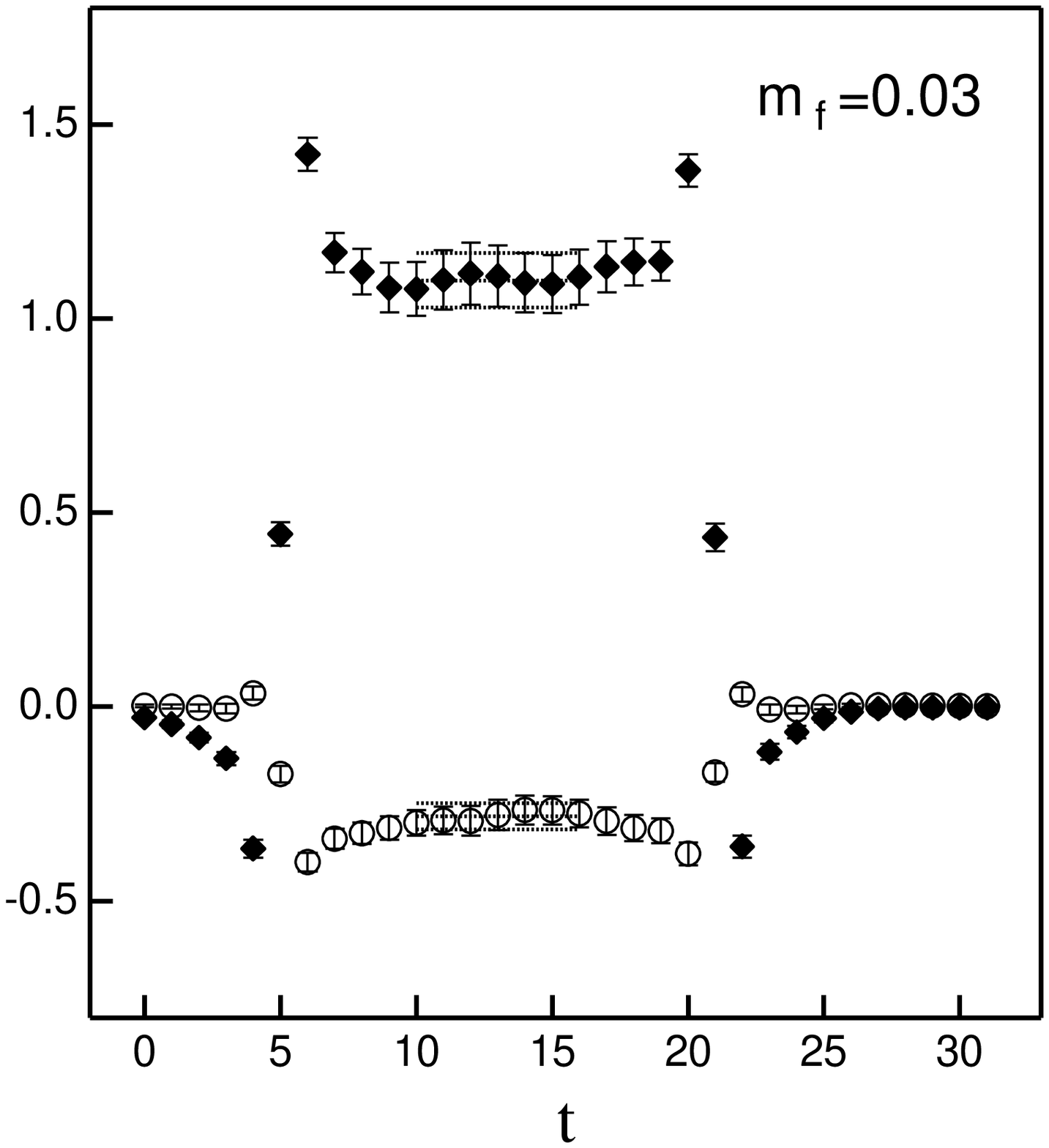}
\linebreak
\vfill\includegraphics[width=2.5in]{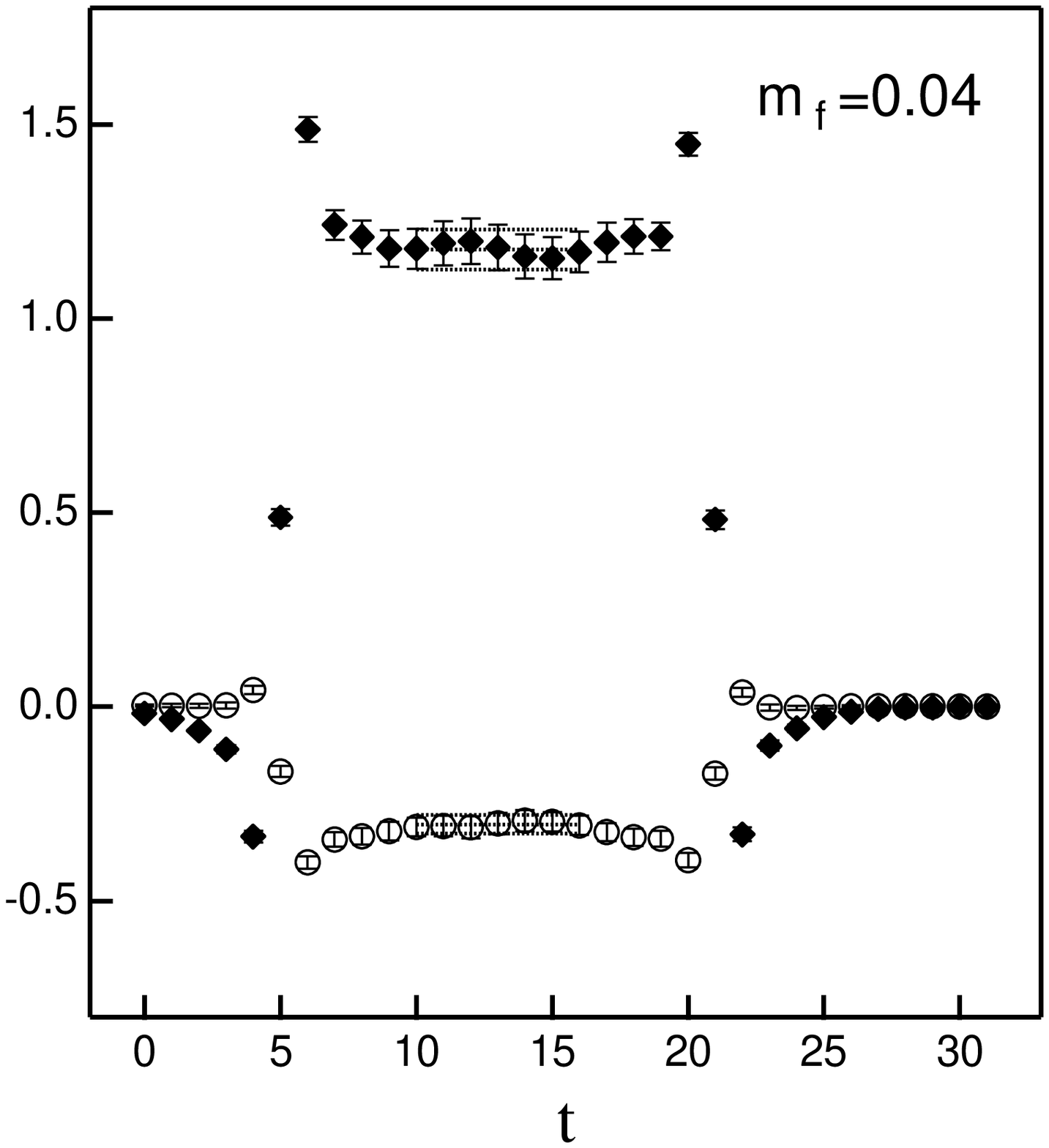}
\includegraphics[width=2.5in]{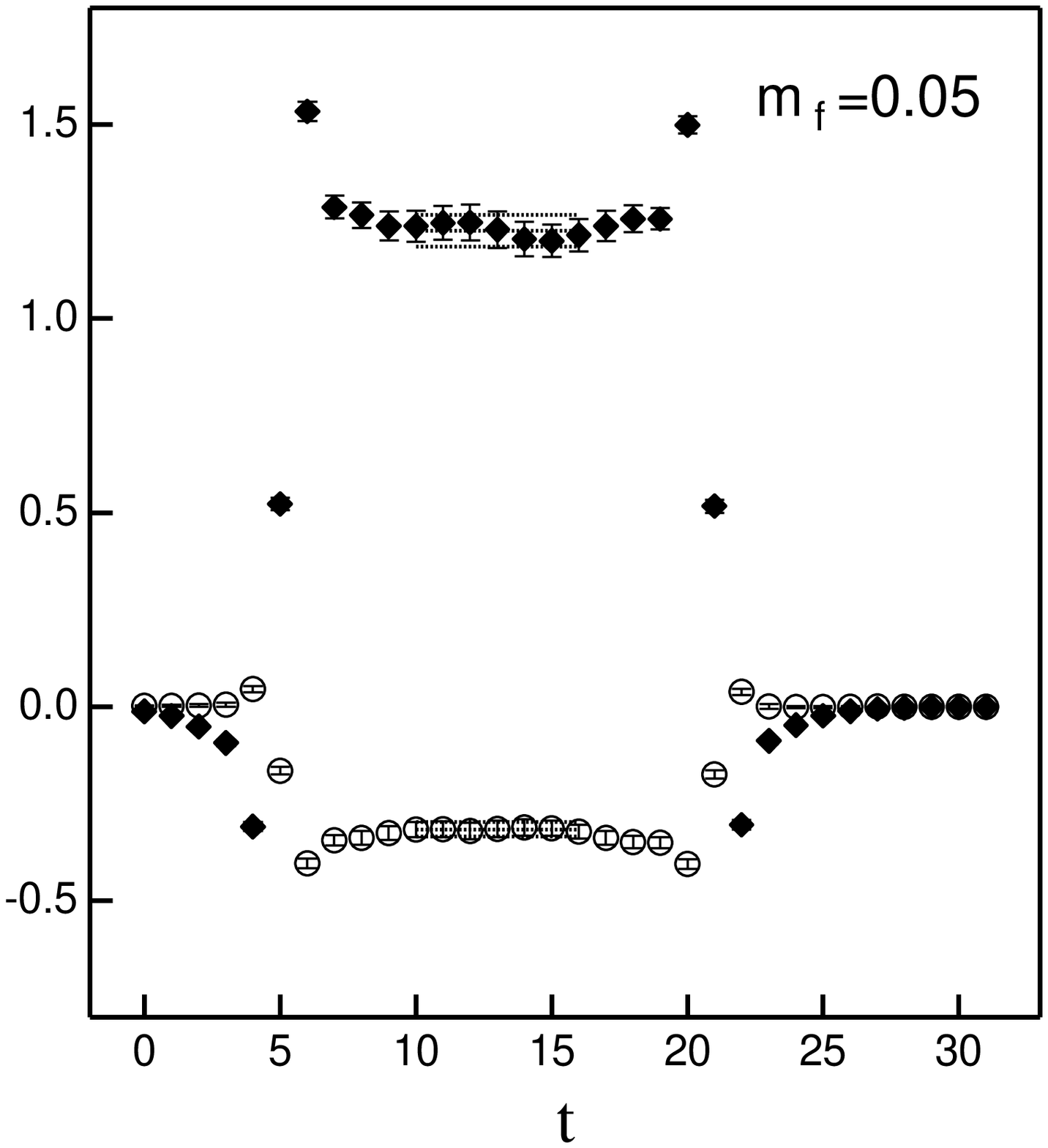}
\linebreak\vfill\end{minipage}
\end{center}
\caption{The lattice estimates of the spin-dependent densities $\Delta u$
(upper curves) and $\Delta d$ (lower curves) for the Wilson gauge
action. Decent plateaus are observed for each quark mass. The
lines denote central values and statistical errors from 
constant fits over the plateaus.}
\label{fig:LDuLDdInsertion_W6}
\end{figure}

\begin{figure}\begin{center}
\includegraphics[width=6in]{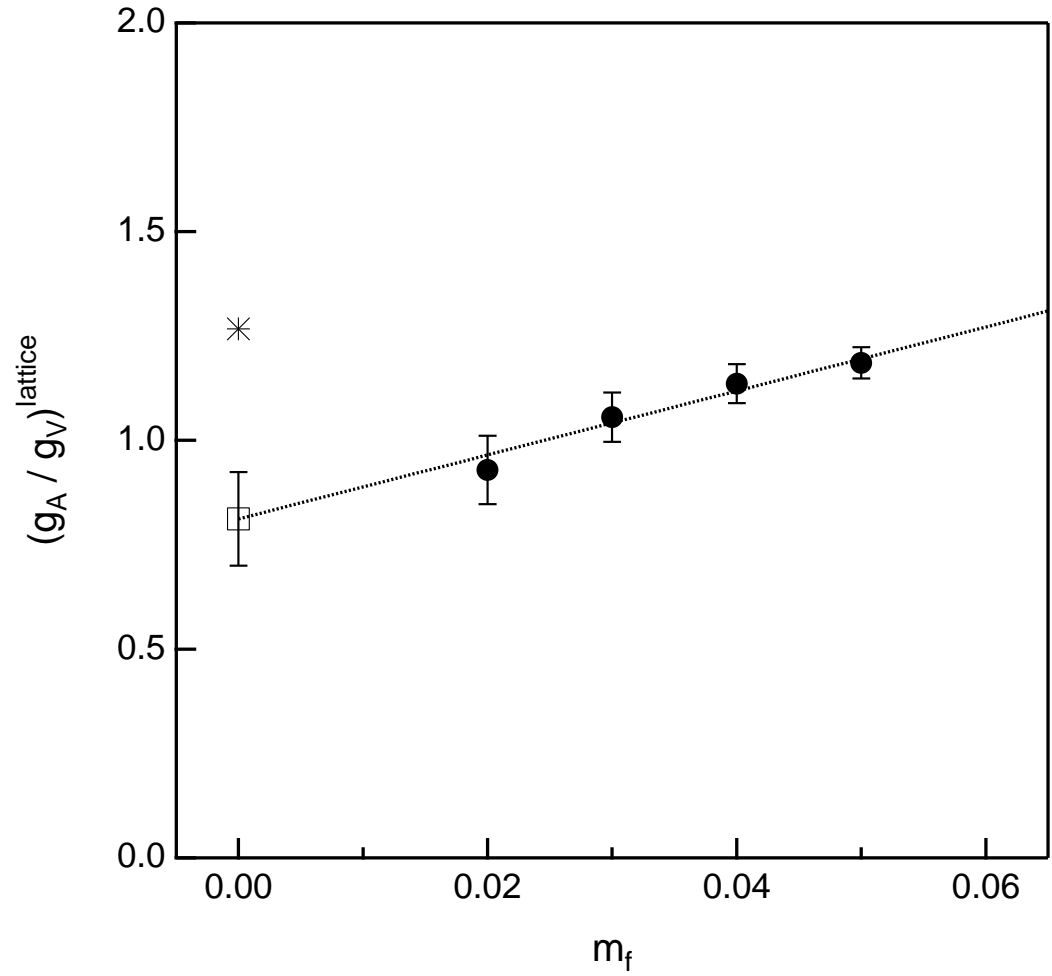}
\end{center}
\caption{The ratio of nucleon charges \(g_{_A}/g_{_V}\), Wilson gauge action,
$V\approx$ (1.6 fm)$^3$. A marked decrease towards the chiral limit is
evident. A 
linear fit significantly underestimates the
experimental value (burst).}
\label{fig:gAgV_W6}
\end{figure}

\begin{figure}
\begin{center}\begin{minipage}[b]{5.25in}
\includegraphics[width=2.5in]{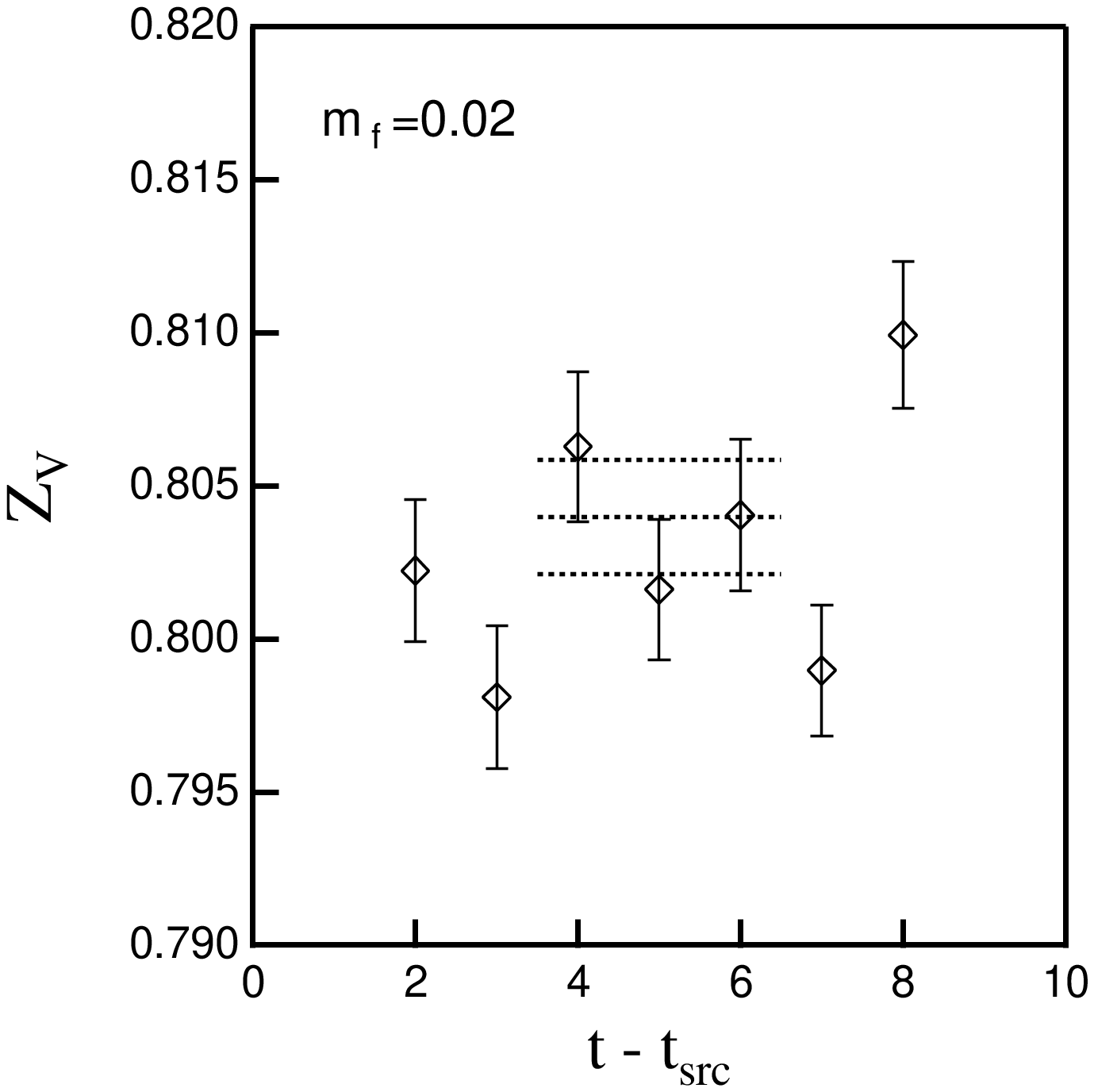}
\includegraphics[width=2.5in]{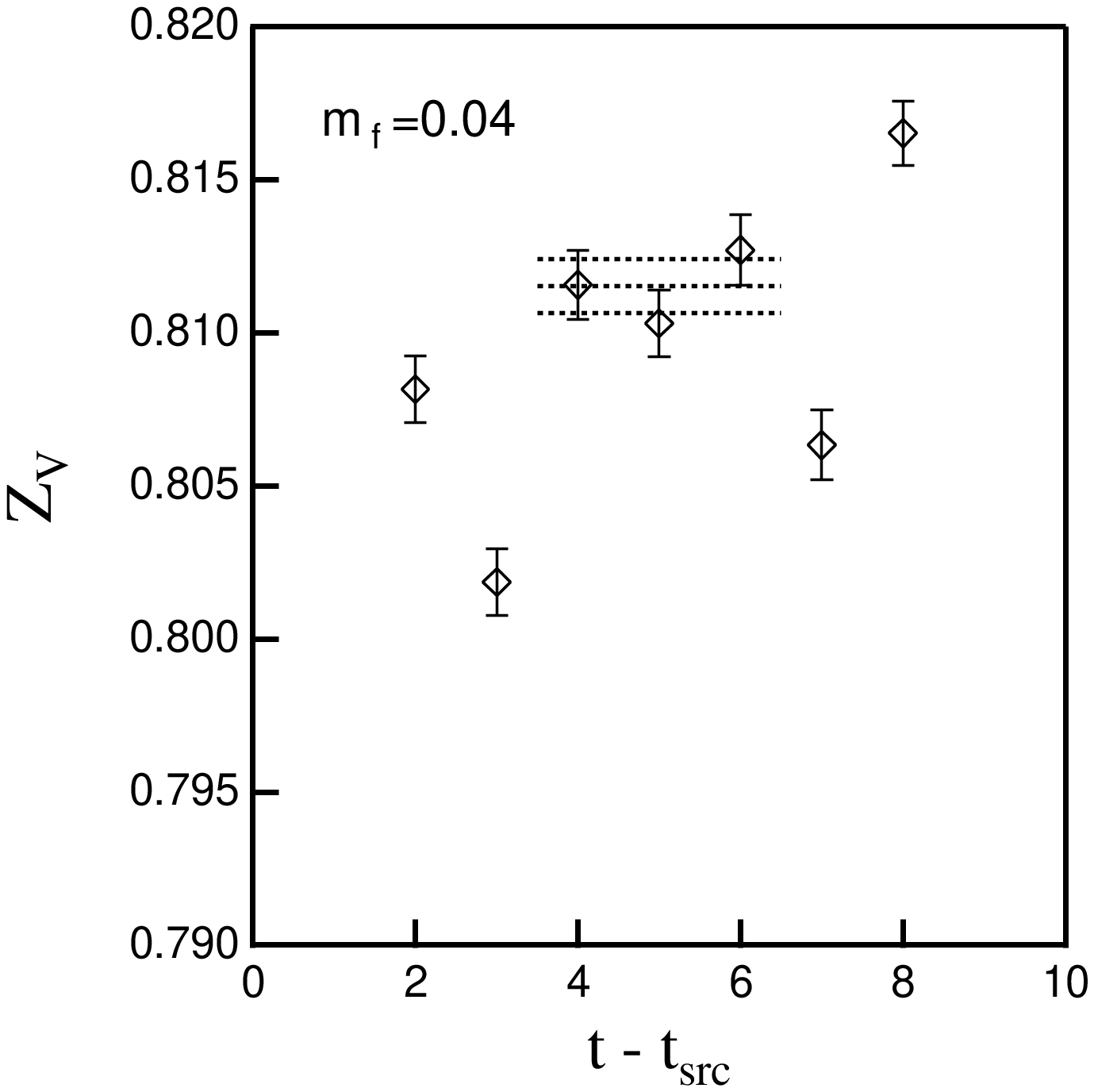}
\linebreak\vfill
\includegraphics[width=2.5in]{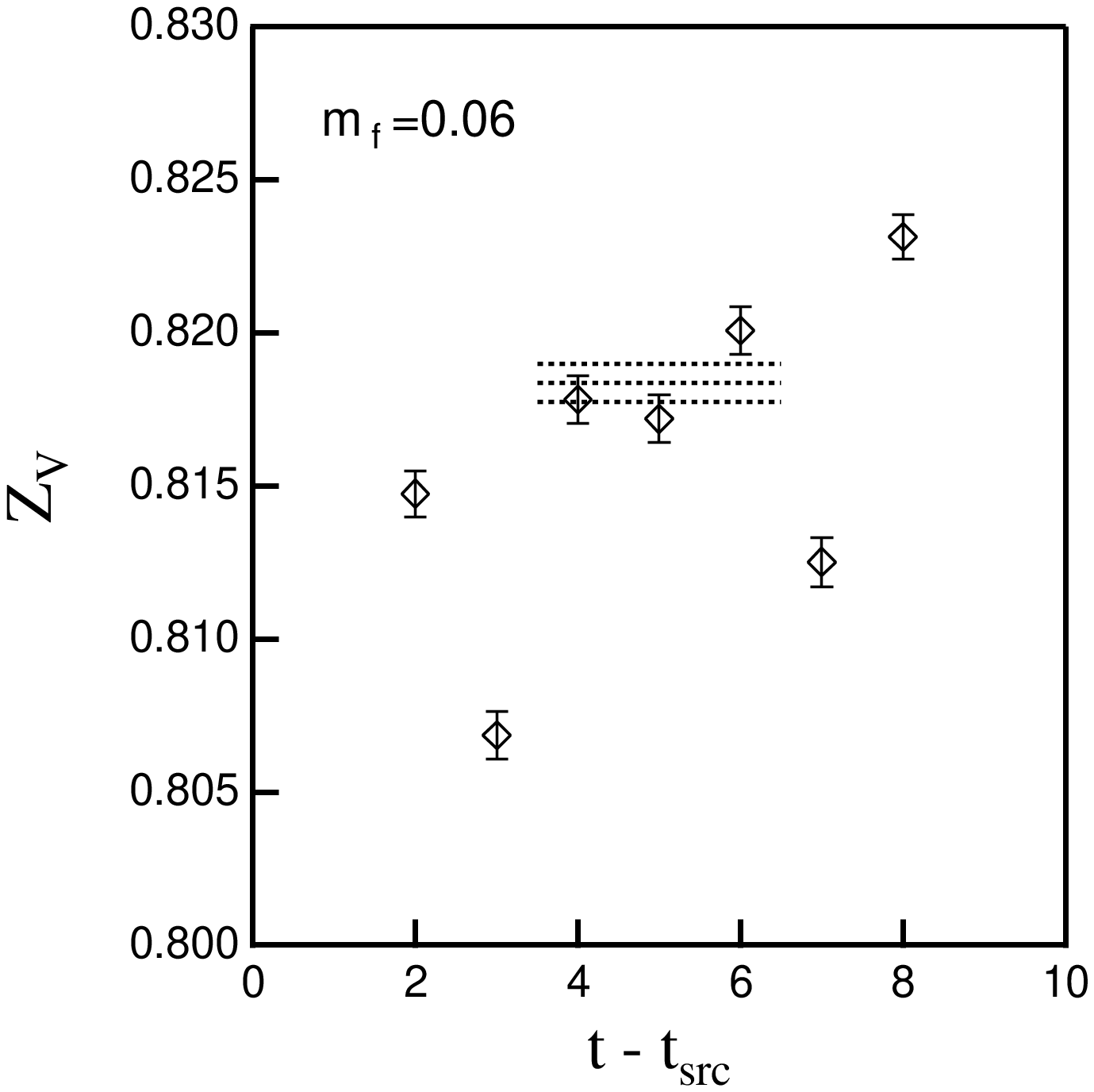}
\includegraphics[width=2.5in]{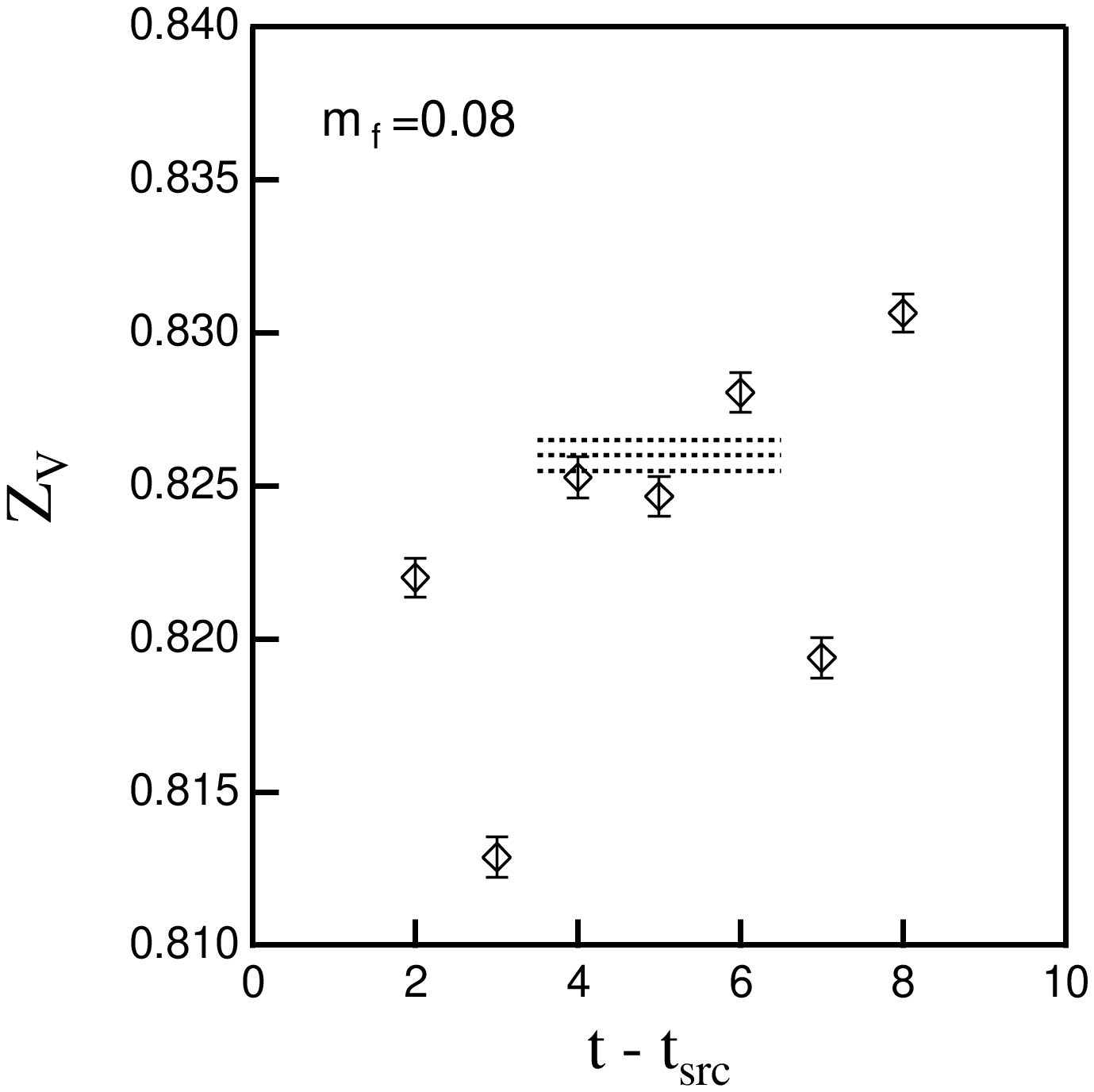}
\linebreak\vfill
\includegraphics[height=2.5in]{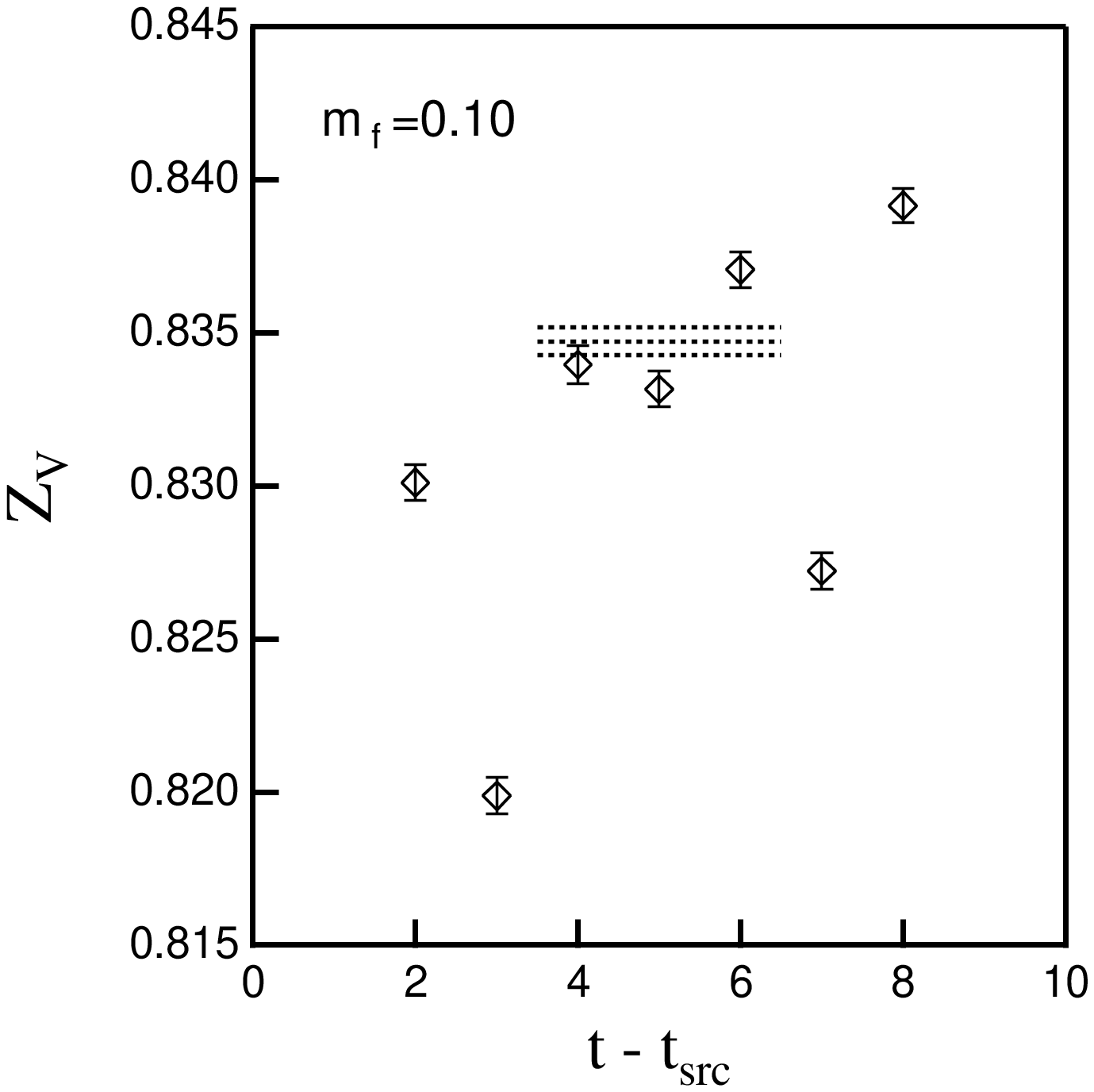}
\end{minipage}
\end{center}
\caption{$\Zv=1/\Gv^{\rm latt}$, DBW2 gauge action, $V\approx$ (2.4 fm)$^3$, as a
function of the current insertion time-slice.  
Note the very fine scale, same as in Figure \ref{fig:ZvInsertion_W6}.
We use the three middle points, the spread of which is less than 0.5 \%.
The lines denote the central values and statistical errors from
constant fits over them.  
}
\label{fig:ZvInsertion_D087}
\end{figure}

\begin{figure}
\begin{center}
\includegraphics[height=5in]{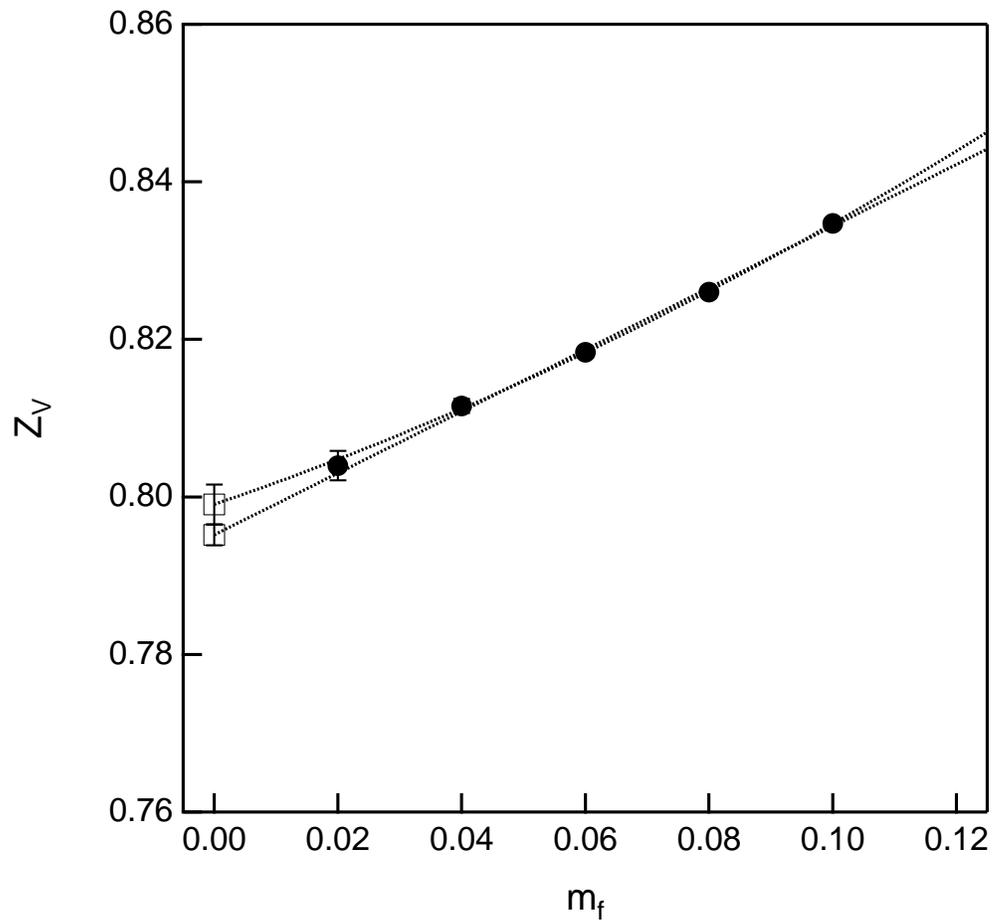}
\end{center}
\caption{Same as Figure~\ref{fig:Zv_mass_dep_W6} except for the DBW2
gauge action, large volume.}
\label{fig:ZAV}
\end{figure}

\begin{figure}
\begin{center}
\includegraphics[height=5in]{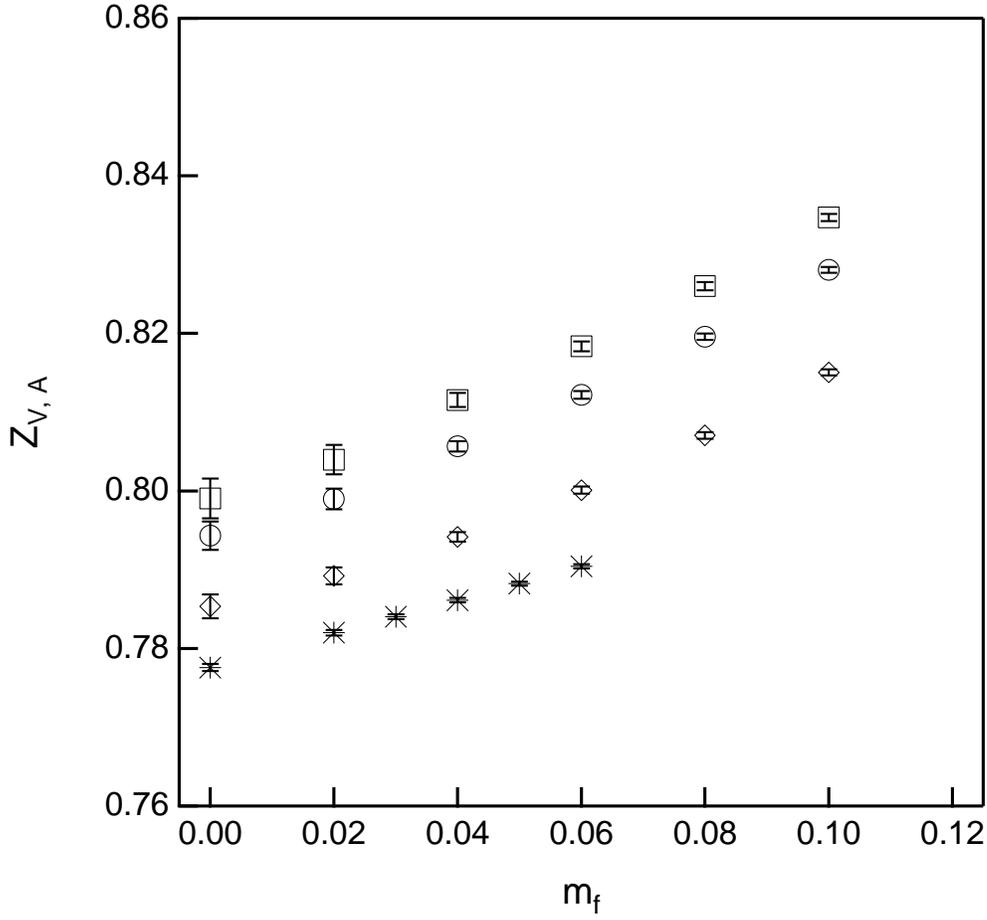}
\end{center}
\caption{Determination of the vector current renormalization from $1/\Gv^{\rm latt}$
(squares), the electromagnetic current (circles), and the d-quark current (diamonds).  The
axial-vector current renormalization (bursts)\cite{Aoki:2002vt} is shown for
comparison. The different renormalization constants differ because of 
${\cal O}(a^2)$ lattice artifacts.}
\label{fig:ZVSYS}
\end{figure}

\begin{figure}
\begin{center}
\begin{minipage}[b]{5.25in}
\includegraphics[width=2.5in]{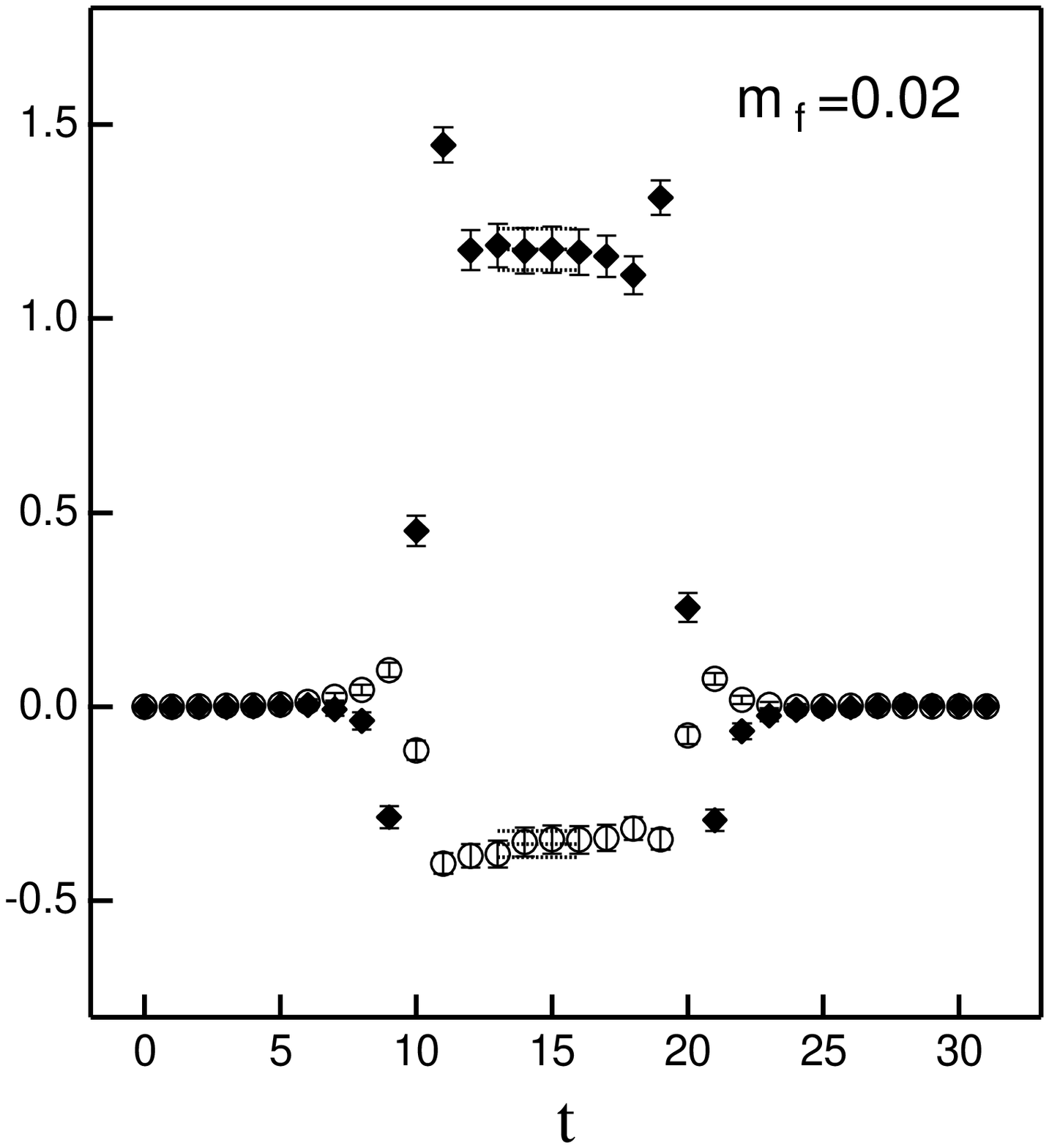}
\includegraphics[width=2.5in]{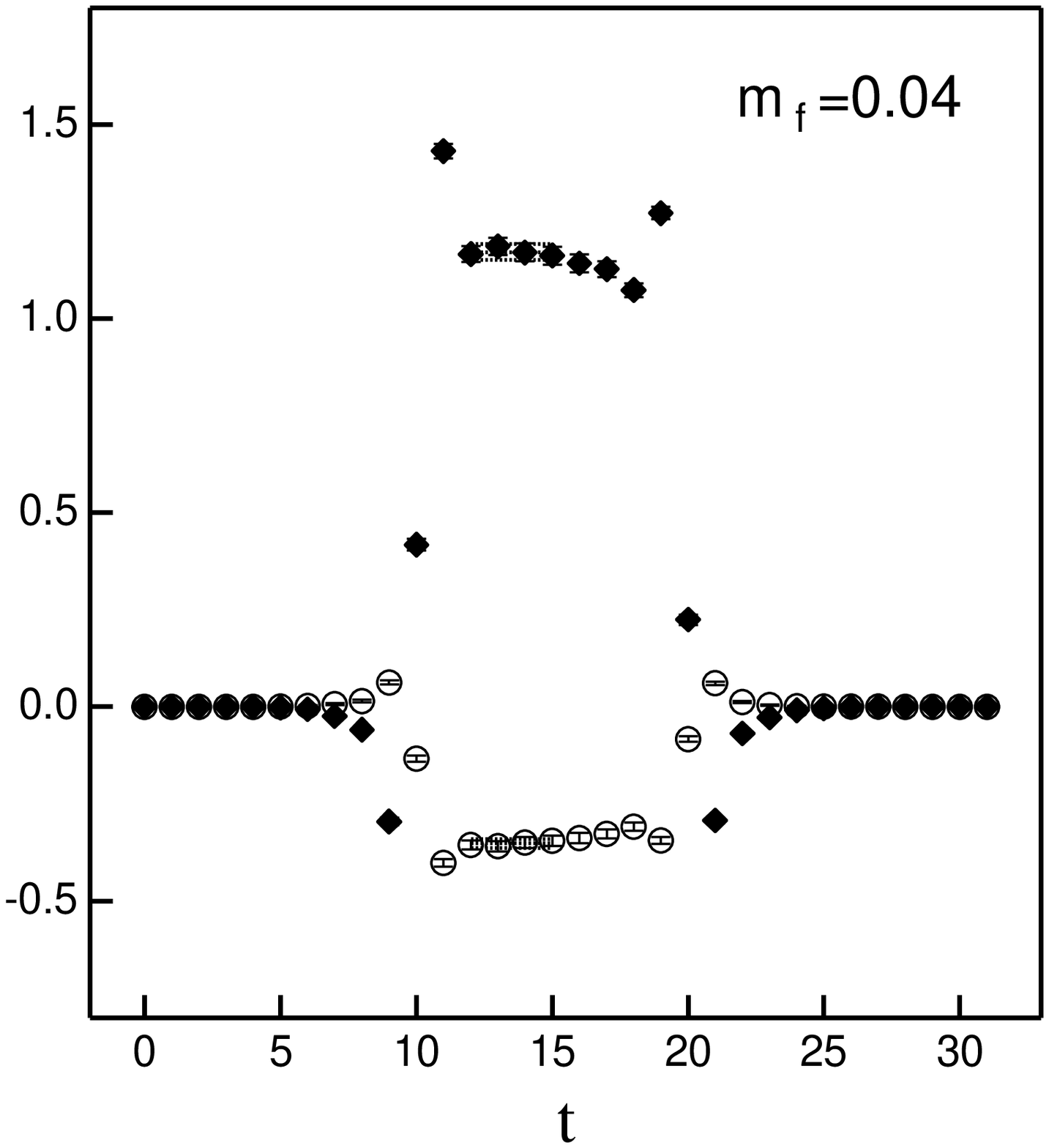}
\linebreak\vfill
\includegraphics[width=2.5in]{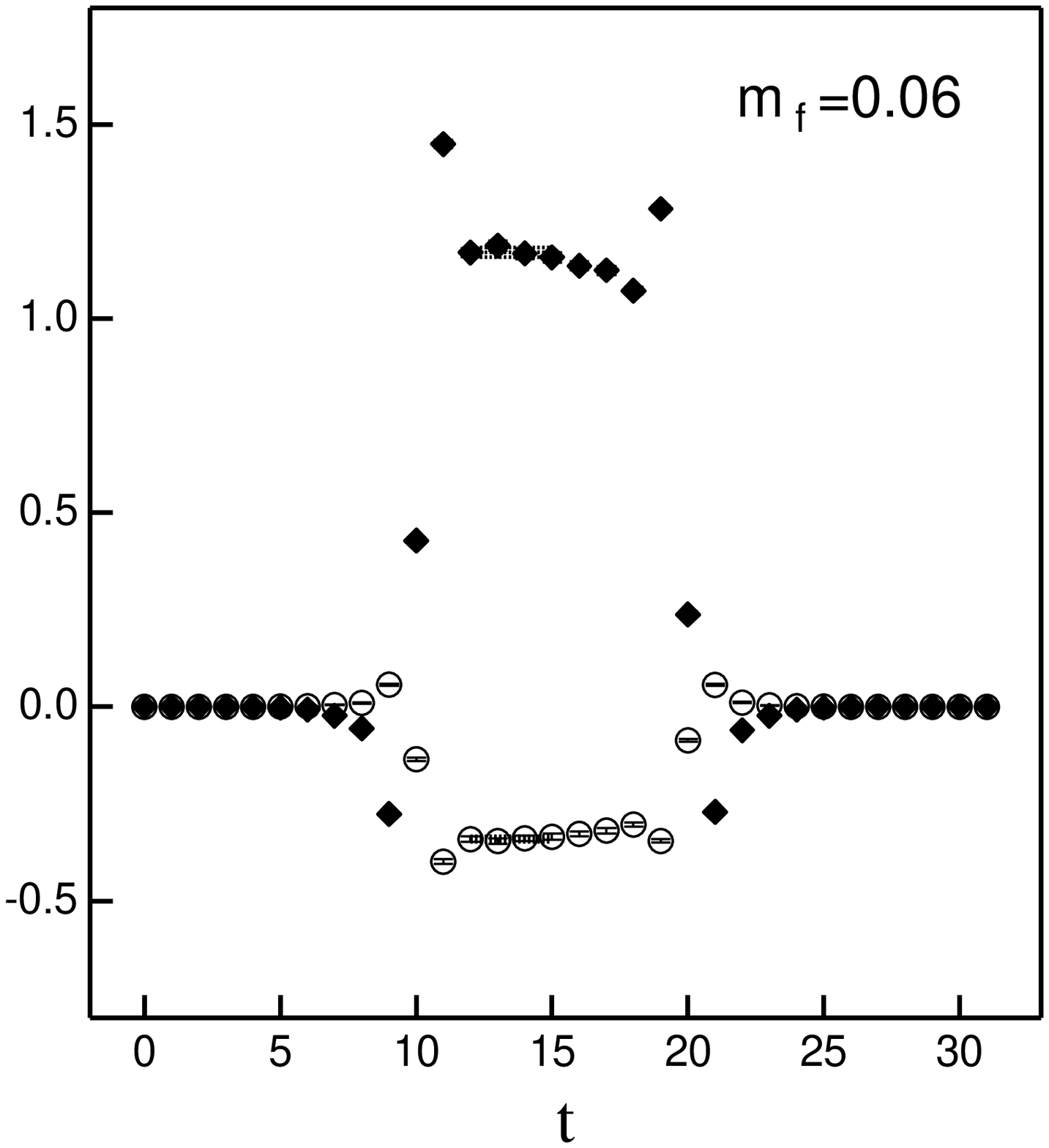}
\includegraphics[width=2.5in]{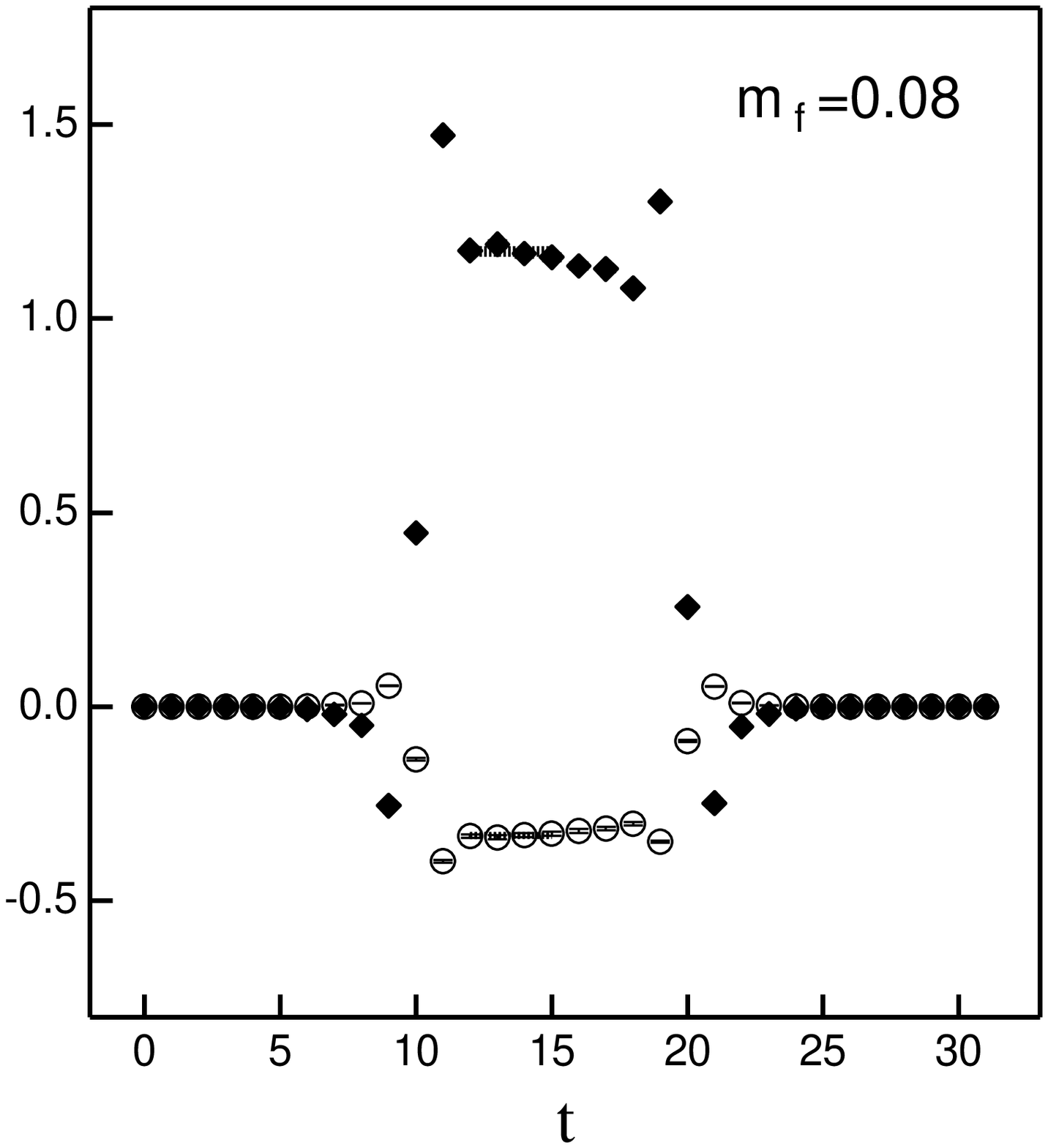}
\linebreak\vfill
\includegraphics[height=2.5in]{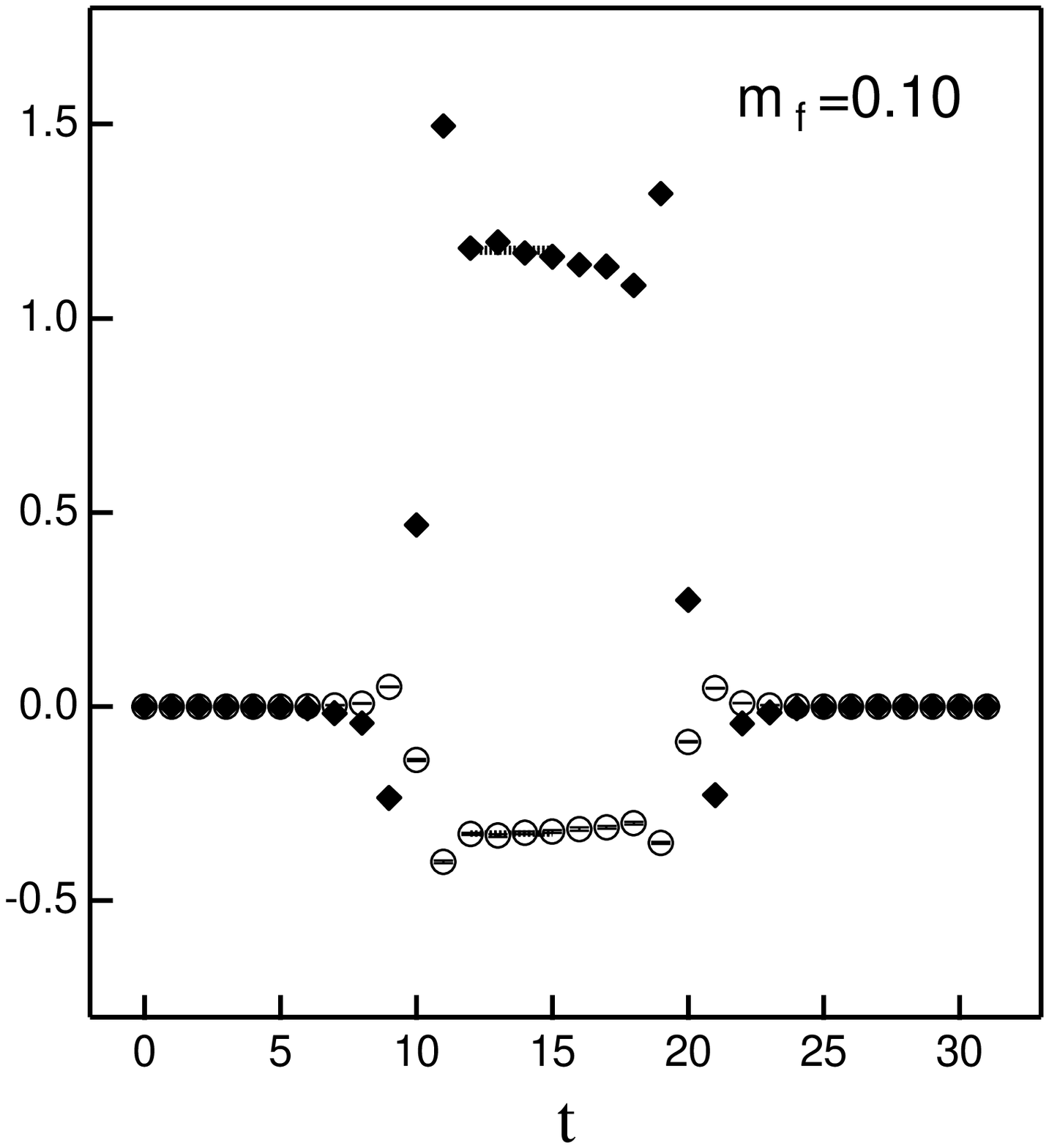}
\end{minipage}
\end{center}
\caption{ $\Delta u$ and $\Delta d$. Same as
Figure~\ref{fig:LDuLDdInsertion_W6}, but for the DBW2 gauge action, sequential
source, and $V=(2.4 {\rm\, fm})^3$.  The plateau is shifted towards the wall
source because the point sink allows more excited state contamination.}
\label{fig:BareDeltaq}
\end{figure}

\begin{figure}
\begin{center}
\includegraphics[height=5.0in]{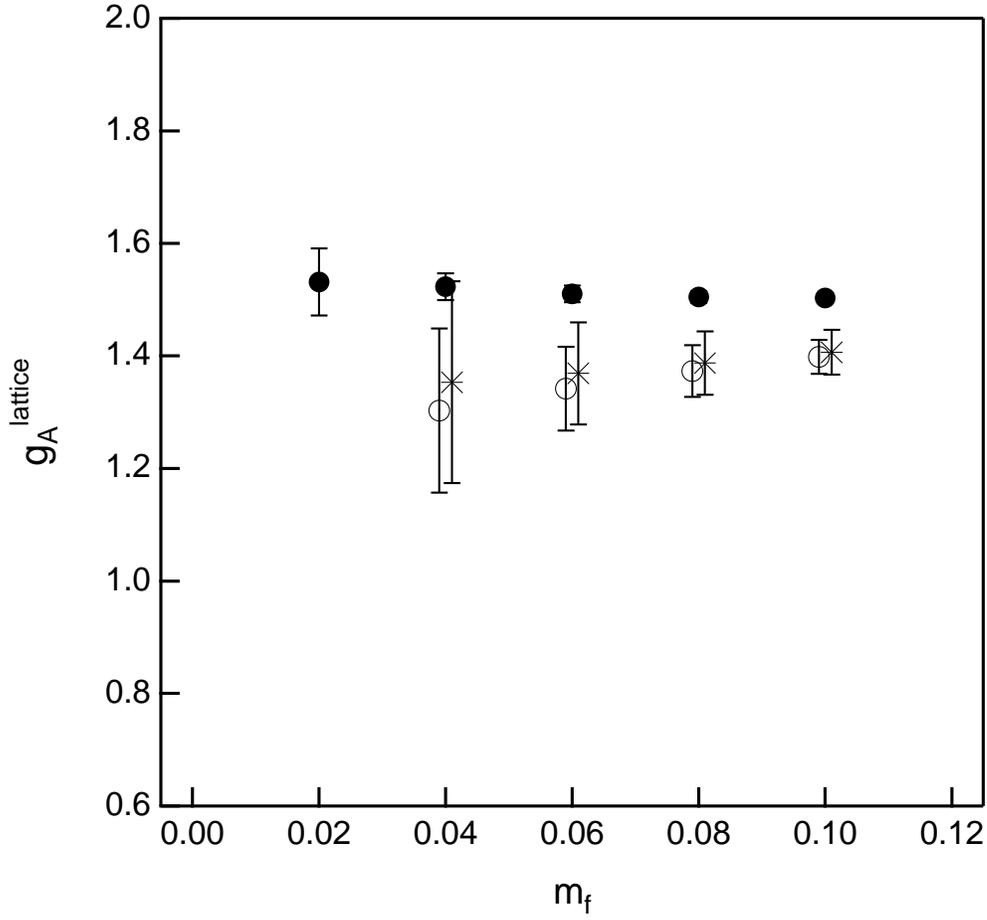}
\end{center}
\caption{Dependence of the unrenormalized nucleon axial charge on quark mass
and lattice volume.  Results from the sequential source
method (circles) and the wall source method (bursts) on the smaller lattice
show good agreement. 
The sequential method provides somewhat smaller %
statistical errors than
the wall method. The larger lattice results (solid circles), obtained with the
sequential method, exhibit higher values than the smaller lattice ones over
the entire range of quark mass studied.}
\label{fig:mfvolgA}
\end{figure}

\begin{figure}
\begin{center}
\includegraphics[height=5.0in]{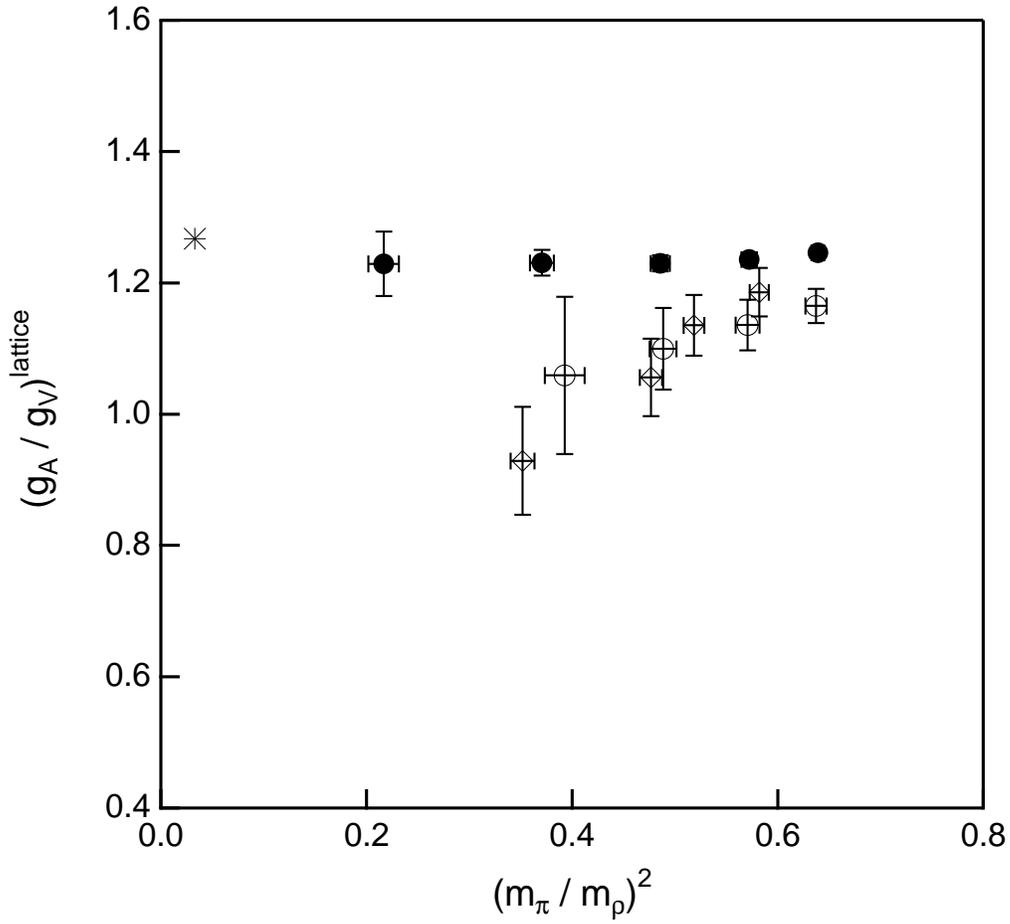}
\end{center}
\caption{The physical ratio of nucleon charges.  DBW2 gauge action results on
two different physical volumes, $(2.4\rm\,fm)^3$ (solid circles) and
$(1.2\rm\,fm)^3$ (open circles), reveal the existence of a significant finite
volume effect. Wilson gauge action results (diamonds), $V\approx(1.6\,\rm fm)^3$,
also appear to be affected by finite volume.}
\label{fig:mfvolgAgV}
\end{figure}

\begin{figure}
\begin{center}
\includegraphics[height=5.0in]{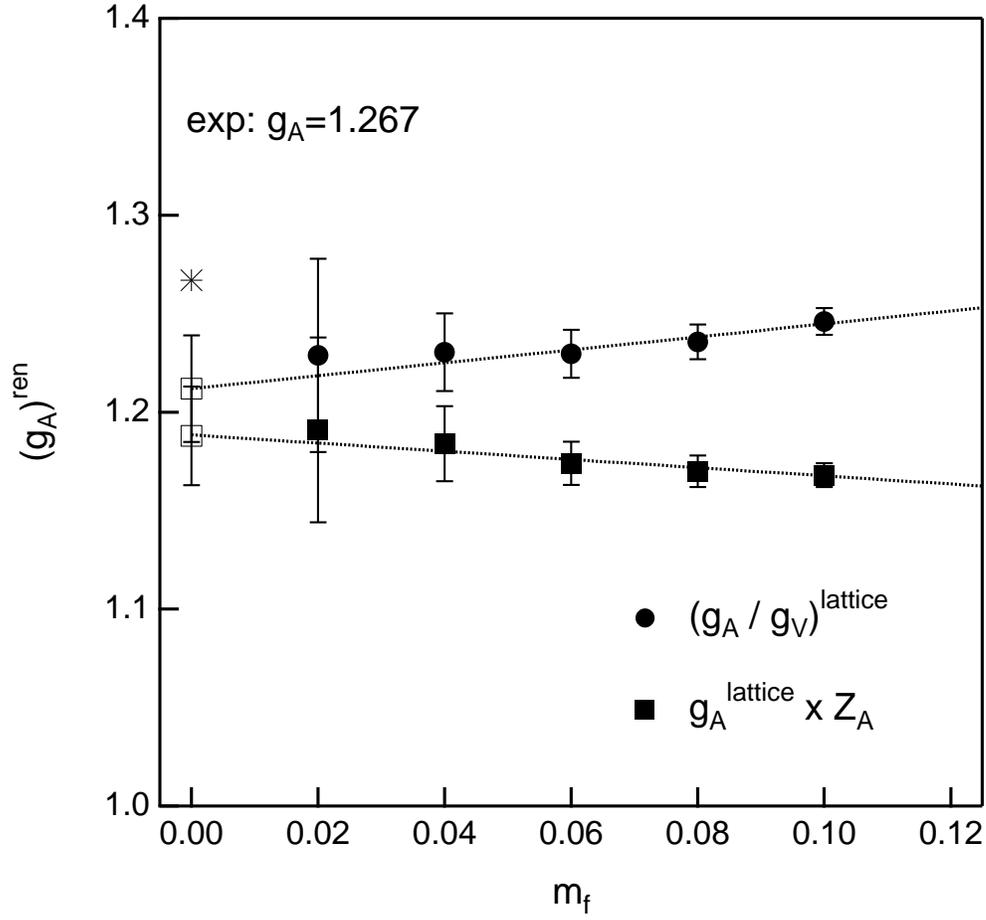}
\end{center}
\caption{Two methods to obtain the physical nucleon axial charge, the ratio of
axial-vector to vector charge (circles), and the lattice axial-vector charge
times the axial-vector current renormalization factor $\Za$ in the chiral
limit from Ref.~\cite{Aoki:2002vt}. They show slightly different quark mass
dependence, but exptrapolate to consistent values. Each underestimates the
experimental value (burst) by rougly five percent.}
\label{fig:AlternategAgV}
\end{figure}

\begin{figure}
\begin{center}
\includegraphics[height=5.0in]{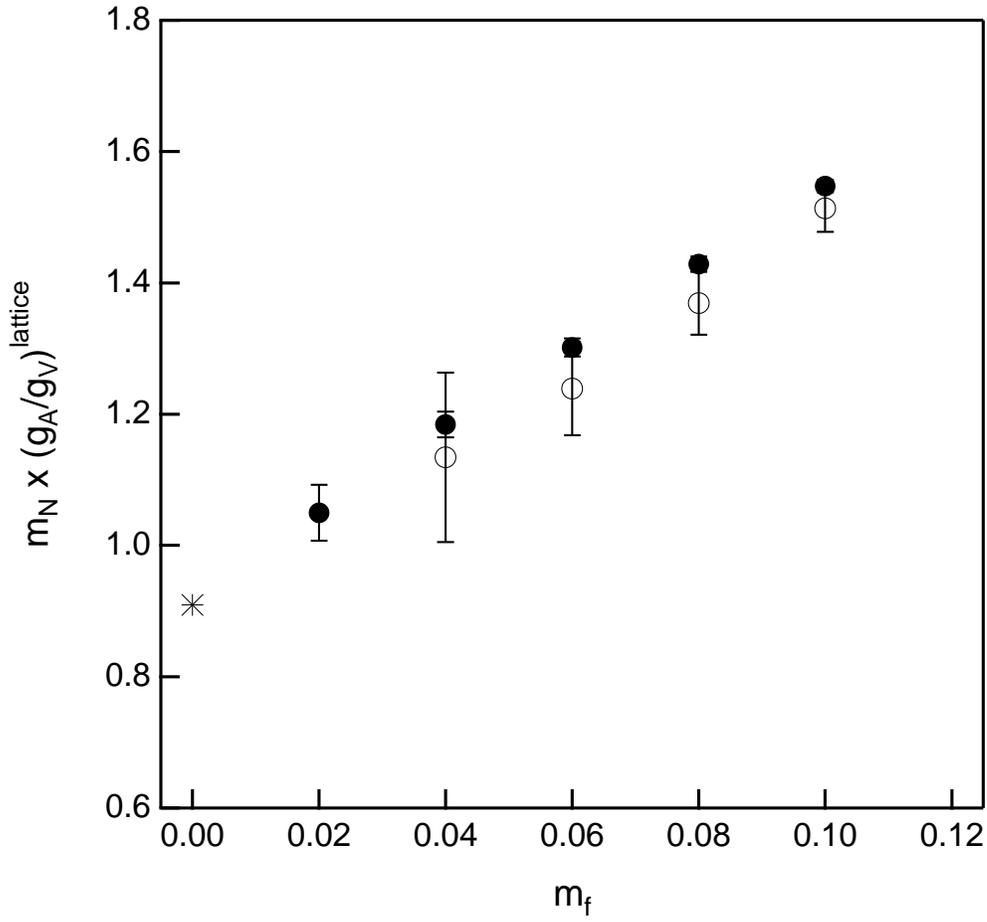}
\end{center}
\caption{The lattice volume and quark mass dependences of the product \(m_{_N}g_{_A}\).    Renormalized.  Closed (large volume) and open (small volume) circles.  All agree within one standard deviation.  No volume dependence is detected.}
\label{fig:mNgAren}
\end{figure}